\documentclass[a4paper,11pt]{article}
\pdfoutput=1
\usepackage{jcappub}
\usepackage{tikz,xcolor,hyperref}
\usepackage{multirow,multicol}
\usepackage{amsmath, amssymb, amsthm, graphicx, epsfig, fancyhdr,epsfig, slashed}
\usepackage[normalem]{ulem}
\usepackage{tikzsymbols}
\usepackage{natbib}
\usepackage{float}
\usepackage{bm}
\usepackage{comment}
\usepackage{graphicx}  
\usepackage{dcolumn} 
\usepackage{bm,relsize}   
\usepackage{slashed}
\usepackage{placeins}
\usepackage{adjustbox}
\usepackage[normalem]{ulem}
\usepackage{lscape}
\usepackage{multirow}
\definecolor{verdes}{cmyk}{0.92,0,0.59,0.4}
\definecolor{rossos}{cmyk}{0,1,1,0.55}
\definecolor{blus}{cmyk}{1,1,0,0.6}
\hypersetup{colorlinks,bookmarksopen,bookmarksnumbered,linkcolor=blus,pdfstartview=FitH,urlcolor=rossos,citecolor=verdes}

\newcommand{\nnn}[1]{{\color{red!50!black}#1}}
\renewcommand{\nnn}[1]{#1}

\def\be{\begin{equation}}
\def\ee{\end{equation}}
\newcommand{\fig}[1]{~\ref{fig:#1}}
\newcommand{\eq}[1]{~{\rm (\ref{eq:#1})}}

 \newcommand{\med}[1]{\langle #1\rangle}
\newcommand{\SU}{\,{\rm SU}}

\newcommand{\U}{\,{\rm U}}

\newcommand{\GeV}{\,{\rm GeV}}
\newcommand{\TeV}{\,{\rm TeV}}

\renewcommand{\refname}{References}

\renewcommand{\eqref}[1]{(\ref{#1})}

\newcommand{\ag}[1]{{\color{orange}[Anish: #1]}}

\definecolor{KBFIred}{RGB}{163,35,47}

\def\beq{\beq\begin{aligned}}
\def\eeq{\end{aligned}\eeq}

\def\beq{\begin{equation}\begin{aligned}}
\def\eeq{\end{aligned}\end{equation}}
\def\bea{\begin{eqnarray}}
\def\eea{\end{eqnarray}}

\makeatletter
\font\ital=cmu10

\def\hhref#1{\href{http://arxiv.org/abs/#1}{arXiv:#1}}
\usepackage{xstring}
\newcommand{\hhrefq}[1]{\IfSubStr{#1}{:}{\href{http://inspirehep.net/search?ln=en&ln=en&p=#1&of=hb&action_search=Search&sf=&so=d&rm=&rg=25&sc=0}{InSpire:#1}}{\hhref{#1}}}

\def\art{\@ifnextchar[{\eart}{\oart}}
\def\eart[#1]#2#3#4#5#6{{\rm #2}, {\em #3 \bf #4} {\rm (#6) #5} ({\em #1})}
\def\article{\@ifnextchar[{\earticle}{\oarticle}}
\def\oarticle#1#2#3#4#5#6{{\rm #1}, {\ital `#6'}, {\rm #2 #3 (#5) #4}}
\def\earticle[#1]#2#3#4#5#6#7{{\rm #2}, {\ital `#7'}, {\rm #3 #4 (#6) #5}  [\hhrefq{#1}]}
\def\hepart[#1]#2{{\rm #2, \sl#1}}
\def\heparticle[#1]#2#3{#2, {\ital `#3'} [\hhrefq{#1}]}
\newcommand{\doi}[1]{\href{http://dx.doi.org/#1}{[link]}}
\def\oarticle#1#2#3#4#5#6{{\rm #1}, {\rm #2 #3 (#5) #4}}
\def\earticle[#1]#2#3#4#5#6#7{{\rm #2}, {\rm #3 #4 (#6) #5}  [\hhrefq{#1}]}
\def\heparticle[#1]#2#3{#2, \hhrefq{#1}}

\newcommand{\hhrefqq}[1]{\IfBeginWith{#1}{10.}{\href{https://doi.org/#1}{doi:#1}}{\hhrefq{#1}}}

\renewenvironment{thebibliography}[1]
{\begin{multicols}{2}[\section*{\refname}]%
		\@mkboth{\MakeUppercase\refname}{\MakeUppercase\refname}%
		\list{\@biblabel{\@arabic\c@enumiv}}%
		{\settowidth\labelwidth{\@biblabel{#1}}%
			\leftmargin\labelwidth
			\advance\leftmargin\labelsep
			\@openbib@code
			\usecounter{enumiv}%
			\let\p@enumiv\@empty
			\renewcommand\theenumiv{\@arabic\c@enumiv}}%
		\sloppy
		\clubpenalty4000
		\@clubpenalty \clubpenalty
		\widowpenalty4000%
		\sfcode`\.\@m}
	{\renewcommand{\@noitemerr}
		{\@latex@warning{Empty `thebibliography' environment}}%
		\endlist\end{multicols}}

 \renewcommand{\sffamily}{\scshape}

\begin{document}
\title{\bf\color{rossos} Gravitational waves and black holes 
from the phase transition 
in models of dynamical symmetry breaking}
\author[a]{\bf Mart\'in Arteaga,}
\author[b]{\bf Anish Ghoshal,}
\author[c]{\bf Alessandro Strumia.}


\affiliation[a]{\em Facultad de Ingenier\'ia, Universidad Privada del Norte, 
Lima, Per\'u
}
\affiliation[b]{\em Institute of Theoretical Physics, Faculty of Physics, University of Warsaw, Poland }
\affiliation[c]{\em Dipartimento di Fisica, Universit\`a di Pisa, Italia}
\emailAdd{martin.arteaga777@gmail.com}
\emailAdd{anish.ghoshal@fuw.edu.pl}
\emailAdd{alessandro.strumia@unipi.it}

\abstract{Theories of dynamical electroweak symmetry breaking
predict a strong first order cosmological phase transition:
we compute the resulting signals,  primordial black holes and gravitational waves.
These theories employ one SM-neutral scalar, plus some extra
model-dependent particle to get the desired quantum potential out of classical scale invariance.
We consider models where the extra particle is a scalar singlet, or vectors of an extended
U(1)  or SU(2) gauge sector.
In models where the extra particle is stable,  it provides a particle Dark Matter candidate with freeze-out abundance
that tends to dominate over primordial black holes.
These can instead be DM in models without a particle DM candidate.
Gravitational waves arise at a level observable in future searches, 
even in regions where DM cannot be directly tested.}

\maketitle


\section{Introduction}
\label{sec:intro}

\medskip
The LIGO-VIRGO-KAGRA interferometers~\cite{Abbott:2016blz}
and Pulsar Timing Arrays~\cite{NANOGrav:2023gor,EPTA:2023fyk,Reardon:2023gzh,Xu:2023wog}
confirmed the expected existence of gravitational waves (GW).
The observed signals seem likely due to astrophysical sources.
The observation of a Stochastic Background of GW (SWGB)
produced during the Big Bang would have a fundamental significance.
SGWB are a unique probe of the early Universe, as the Universe is transparent to GWs right from the wee moments of the Big Bang, unlike other cosmic relics like photons and neutrinos. 
Although LIGO-VIRGO only set an upper limit on SGWB~\cite{LIGOScientific:2016jlg, LIGOScientific:2019vic, KAGRA:2021kbb}, 
increased sensitivity in the nHz-kHz frequency range should be reached by possible future observations such as
SKA~\cite{Weltman:2018zrl}, GAIA/THEIA~\cite{Garcia-Bellido:2021zgu}, MAGIS~\cite{MAGIS-100:2021etm}, AION~\cite{Badurina:2019hst}, AEDGE~\cite{AEDGE:2019nxb}, $\mu$ARES~\cite{Sesana:2019vho}, LISA~\cite{LISA:2017pwj}, TianQin~\cite{TianQin:2015yph}, Taiji~\cite{Ruan:2018tsw}, DECIGO~\cite{Kawamura:2020pcg}, BBO~\cite{Corbin:2005ny}, ET~\cite{Punturo:2010zz},  CE~\cite{Reitze:2019iox}.
GW searches at MHz-GHz higher frequencies 
are discussed~\cite{Aggarwal:2020olq, Berlin:2021txa, Herman:2022fau, Bringmann:2023gba}.

\smallskip

Among various cosmological mechanisms for producing  a SGWB~\cite{Roshan:2024qnv}, cosmological strong first-order phase transitions (FOPTs)
(see e.g.~\cite{Caprini:2015zlo}) offer plausible beyond the Standard Model (BSM) signals, up to high scales. 
Indeed the SM predicts two (electroweak and QCD) phase transitions, none of which is of first-order~\cite{Kajantie:1996mn, Bhattacharya:2014ara}. 
Therefore, the detection of a GW signal compatible with a FOPT would be evidence of BSM physics. 
FOPTs develop by the formation of bubbles that expand, collide and percolate. The violent collisions between the bubble walls (and the motion of the surrounding thermal plasma) lead to the production of stochastic GWs.
The typical frequency of such gravitational waves is the red-shifted Hubble rate during the phase transition, 
$f_{\rm peak}\sim T_0 T/M_{\rm Pl} $,
\nnn{where $T_0$ is the current CMB temperature}. 
A FOPT around the electroweak scale, $T\sim 100$ GeV, would peak at $f\sim$ mHz~\cite{Grojean:2006bp} which is in the frequency sensitivity band of space-based GW experiments such as LISA~\cite{LISA:2017pwj}, 
whereas ground-based experiments such as LIGO-VIRGO~\cite{LIGOScientific:2014pky, VIRGO:2014yos} and ET~\cite{Punturo:2010zz} around 100 Hz  
can probe FOPTs up to $T\sim 10^8$ GeV~\cite{Dev:2016feu}, 
 beyond the reach of collider experiments.  

A detectable GW signal arises  if the phase transition is strongly first-order. 
This  is fulfilled by approximately scale-invariant theories~\cite{Meissner:2006zh}, 
where scale symmetry is broken dynamically through the Coleman-Weinberg mechanism~\cite{Coleman:1973jx} 
after a significant  amount of supercooling~\cite{Witten:1980ez}. 
So, bubble collisions take place in the vacuum, enhancing the amplitude of the corresponding GW signal~\cite{DelleRose:2019pgi,VonHarling:2019rgb,Ghoshal:2020vud}.\footnote{See e.g.~\cite{Jaeckel:2016jlh, Jinno:2016knw, Marzola:2017jzl, Iso:2017uuu, Chao:2017ilw,1809.01198, Prokopec:2018tnq, Brdar:2018num, Marzo:2018nov, Hasegawa:2019amx, Ellis:2020nnr,  Chikkaballi:2023cce, Ahriche:2023jdq} for explorations of this mechanism in other BSM contexts. This has also been used in the context of baryogenesis via leptogenesis~\cite{Huang:2022vkf,Dasgupta:2022isg,Borah:2022cdx}, complementarity with collider searches~\cite{Dasgupta:2023zrh} and of the generation of the Planck scale~\cite{Ghoshal:2022qxk}.}

\medskip

In this work we analyze the cosmological implications of such BSM phase transitions. 
Besides an enhanced GW signal, a supercooled FOPT also leads to the formation of primordial black holes (PBHs) via different mechanisms.
Bubble collisions have been studied e.g. in~\cite{Khlopov}. 
We focus on a mechanism suggested in~\cite{Sato:1981bf,Kodama:1981gu,Maeda:1981gw,Sato:1981gv,Hawking:1982ga,Kodama:1982sf}
that recently received interest~\cite{Lewicki:2019gmv,Ashoorioon:2020hln,Kawana:2021tde, Liu:2021svg,Jung:2021mku,Hashino:2022tcs,Huang:2022him,Kawana:2022lba,Kawana:2022olo,Kierkla:2022odc,Kierkla:2023von,Hashino:2021qoq,He:2022amv,Gehrman:2023esa,Lewicki:2023ioy,Gouttenoire:2023naa,Gouttenoire:2023pxh,2307.11639,Salvio:2023ynn,2311.03406,Goncalves:2024vkj,Conaci:2024tlc,2404.06506,2404.06506}. 
FOPTs proceed via the nucleation of bubbles of the broken phase in an initial background of the symmetric phase~\cite{Coleman:1977py,Callan:1977pt,Linde:1981zj}. 
In a supercooled regime, the energy density of the Universe in the symmetric phase is dominated by the vacuum energy which acts as a cosmological constant and leads to an inflationary period. In a nucleated bubble, instead, such energy is quickly converted into radiation and the corresponding patch expands slower.
Since bubble nucleation is a stochastic phenomenon, and since, in a supercooled regime, regions outside the nucleated bubbles dilute faster than those inside, 
a region where nucleation happens later becomes over-dense and eventually collapses, if large enough, into a PBH. 

The paper is organised as follows.
In section~\ref{sec:form} we present the basic framework and summarise standard results about how gravitational waves
and Primordial Black Holes  are later computed in specific models: 
\begin{itemize}
\item  In section~\ref{sec:mods'} we consider a minimal model that realises dynamical electroweak symmetry breaking 
 by adding two scalars to the SM: a singlet $s$ that acquires a vacuum expectation value, and
 another singlet $s'$ (a possible DM candidate) with couplings that induce the needed quantum effects.
 
 \item In section~\ref{sec:modSU2} we consider a model where $s$ is doublet under an extra $\SU(2)$ group;
its vectors are stable and provide DM candidates. 

\item In section~\ref{sec:modU1} we consider a model where $s$ is charged under a U(1) gauge group,
that we identify as U(1)$_{B-L}$ such that there is no particle DM.
Only Primordial Black Holes provide viable Dark Matter candidates.

\end{itemize}
These different models provide similar signals.
Conclusions and a summary of results are given in section~\ref{concl}.

\section{Formalism for dynamical electroweak symmetry breaking}\label{sec:form}
Dimension-less scalar potentials with quartic couplings that run to negative values when renormalized down to low energy lead to dynamical symmetry breaking~\cite{Coleman:1973jx}
via a first-order phase transition.
This mechanism can induce the breaking of the electroweak symmetry.
The minimal implementation with just the Standard Model Higgs doublet $H$
is excluded because it predicts a too light Higgs boson, $m_h \approx 7\GeV$~\cite{Gildener:1976ih}.
This wrong prediction is avoided by adding an extra scalar $s$ neutral under the SM gauge group.
The tree-level scalar potentials have the form
\beq\label{eq:V0}
V_{\rm tree}(H,s) = V_\Lambda + \lambda_H|H|^4+\frac{\lambda_S}{4} s^4+\frac{\lambda_{H S}}{2}|H|^2 s^2  +\cdots
\eeq
where $\cdots$ denotes possible extra scalars.
If the beta function of $\lambda_S$ is positive, $\beta_{\lambda_S} =  d\lambda_S/d\ln\mu > 0$, 
the loop potential
\beq V_{\rm loop}(0,s) \simeq  \lambda_S(s) \frac{s^4}{4}\simeq
\frac{\beta_{\lambda_S}}{2}  \, \frac{s^4}{4} \, \ln \frac{s^2}{w^2 e^{1/2}}
\eeq
has a minimum at $\med{s}=w$, where $w$ effectively is a free parameter.
The $s$ mass is loop suppressed, $m_s = w \sqrt{\beta_{\lambda_S}}$.
The dimensionful constant $V_\Lambda \approx \beta_{\lambda_S} w^4/16$ needs to be added to eq.\eq{V0} such that the loop potential vanishes
at its minimum, as required by the tiny observed cosmological constant.
A small negative value of $\lambda_{HS}<0$ then leads to electroweak dynamical symmetry breaking as
\beq H = \frac{1}{\sqrt{2}}
\begin{pmatrix}
0\cr h
\end{pmatrix},\qquad \med{h}=v \simeq w \sqrt{\frac{-\lambda_{HS}}{2\lambda_H}} 
\eeq
in the unitary gauge.
Here $\lambda_H \approx 0.126$ is the SM Higgs quartic, up to small corrections.
During the Big Bang, $s$ acquires a thermal mass such that the universe initially remains trapped in the false vacuum at $s=h=0$.
This super-cooling starts at the temperature $T_{\rm eq}$ such that the potential energy 
becomes more important than the thermal energy
\beq\label{eq:Teq}
\frac{\pi^2 g_* T_{\rm eq}^4}{30} = V_\Lambda\eeq
with $g_* \approx 106.75$ the number of SM degrees of freedom.
Next the temperature and thereby the thermal mass drops, and
super-cooling can end via a first order phase transition that happens when nucleation becomes faster than the expansion rate.
At this point the energy $V_\Lambda$ stored in the potential gets released.
We will focus on large enough couplings such that reheating is fast, so that the universe reheats up to temperature
\beq \label{eq:TRH}
T_{\rm RH} \simeq T_{\rm eq} \simeq w \left(\frac{15 \beta_{\lambda_{S}}}{8\pi^2 g_*}\right)^{1/4}. \eeq
Extra particles are needed to get the desired $\beta_{\lambda_S}>0$ running of $\lambda_S$,
leading to a variety of models of dynamical electroweak symmetry breaking.
Three models will be described and computed in sections~\ref{sec:mods'}, \ref{sec:modSU2}, \ref{sec:modU1}.
In this section we describe the model-independent common formalisms.


\subsection{Computing the thermal potential}
At finite temperature $T$, the potential of generic scalars $\phi$
receives an additive thermal contribution $V_{T}$ given at leading order by the standard expression
\begin{equation}
	V_T(\phi, T)=\frac{T^4}{2 \pi^2} \sum_{\mathrm{b}} J_B\left(\frac{m_{\mathrm{b}}^2(\phi)}{T^2}\right)+\frac{T^4}{2 \pi^2} \sum_{\mathrm{f}} J_F\left(\frac{m_{\mathrm{f}}^2(\phi)}{T^2}\right)
\end{equation}
where `b' and `f' stand for all {bosons} and {fermions} present in the theory. 
The sums run over all field-dependent masses $m^{2}_{{\rm b},{\rm f}}(\phi)$.
The thermal integrals $J_B$ and $J_F$ are 
\begin{equation}
	J_{B / F}\left(y^2\right)= \int_0^{\infty} dx\, x^2 \ln \left[1 \mp \exp \left(-\sqrt{x^2+y^2}\right)\right].
\end{equation}
These functions have the small-field expansion
\beq 
\begin{array}{ll}\displaystyle
J_B(y^2)\simeq - \frac{\pi^4}{45} + \frac{\pi^2}{12} y^2-  \frac{\pi}{6}  y^3 + \frac{y^4}{32} \ln\frac{y^2}{a_B} +\cdots,\\
\displaystyle
J_F(x^2) \simeq \frac{7\pi^4}{360} - \frac{\pi^2}{24}y^2 - \frac{y^4}{32}\ln\frac{y^2}{a_F}+\cdots
\end{array}
\eeq
with $a_B =(4\pi)^2 e^{3/2-2-\gamma_E} = 16 a_F$.
The quadratic term dominates at small field values, so that
positive squared thermal masses trap scalars at the local minimum $\med{s}=\med{h}=0$ of $V_T$.
For example, the SM contribution to the Higgs thermal mass is
\beq
M_{hT}^{2} =\bigg(\frac{3}{16} g^2_2 + \frac{1}{16}
g^2_Y + \frac{1}{4} y^2_t + \frac12 \lambda_H\bigg)T^2 .
\label{eq:MhT}
\eeq
The thermal mass of $s$ arises from its model-dependent couplings needed to have the running $\lambda_{\beta_S}>0$.
The quartic terms in $J_{B,F}$ imply that the quartic couplings get renormalised around $T$.
The bosonic function $J_B$ also contains to a cubic term, relevant in tunnelling computations.

\subsection{Computing the phase transition}
Following the standard tunnelling formalism,
the space-time density of bubble nucleation rate at finite temperature is 
\be\label{eq:Gamma_rate_S3T}
 \gamma \approx  T^{4}\left(\frac{S_{3}}{2\pi T}\right)^{3/2}\, e^{-S_3/T} 
\ee
where $S_3$ is the action of the dominant O(3)-invariant thermal bounce.
In our computations the tunnelling action $S_3$ is numerically computed from the thermal potential,
that includes the full $J_{B,F}$ functions, plus
higher-order `daisy' contributions.

The following analytic approximation allows to understand the main features.
The finite-temperature potential along the $s$ direction can be approximated as 
\beq \label{eq:VTapprox}
V \simeq \frac{m^2}{2} s^2 - \frac{k}{3} s^3 - \frac{\lambda}{4}   s^4 + \cdots\eeq
with roughly constant $\lambda = - \lambda_S(T) \simeq \beta_{\lambda_S} \ln(w/T)>0$.
This coupling runs larger at lower $T$, facilitating the phase transition.
Indeed the bounce action corresponding to the potential of eq.\eq{VTapprox} can be approximated as (see e.g.~\cite{Salvio:2023ynn})
\beq S_3 \approx \frac{27\pi m^3}{2 k^2}\frac{1+e^{-\tilde\lambda^{-1/2}}}{1+9\tilde{\lambda}/2},\qquad
\tilde\lambda = \lambda \frac{m^2}{k^2}.
 \eeq
In the limit of negligible cubic coupling, $k=0$, the action reduces to $S_3 \simeq 6\pi  m/\lambda$.
After crossing $\lambda=0$, the running coupling $\lambda$ becomes larger at lower temperature.
So $S_3/T$ tends to increase at lower $T$ proportionally to $1/\ln(w/T)$.
Super-cooling is ended by nucleation at low enough $T_{\rm nuc}$
provided that large enough couplings make tunnelling faster than the Hubble rate.
The nucleation temperature $T_{\rm nuc}$ is defined as
\begin{equation}\label{eq:Gamma_H4}
\gamma(T_{\rm nuc}) \approx H^{4}(T_{\rm nuc}).
\end{equation}
The Hubble rate is approximated as
\begin{equation}\label{eq:H2_supercooling}
	H^{2} = \frac{\rho_{\rm rad} + V_{\Lambda}}{3 \bar{M}_{\text{Pl}}^{2}}  ,\qquad \rho_{\rm rad} = \frac{\pi^{2} }{30} g_{*} T^{4}
\end{equation}
with $\rho_{\rm rad} \ll V_\Lambda$ during super-cooling.
As a consequence the relative amount of released energy in a super-cooled phase transition is large $\alpha \approx V_\Lambda /\rho_{\rm rad}\gg 1$,
giving large gravitational wave signals.\footnote{The parameter $\alpha$, quantifying the strength of the transition, varies in definition across the literature, with primary interpretations focusing on latent heat and the trace anomaly~\cite{Helmboldt:2019pan}. 
It is generally defined as either the ratio of the transition's latent heat to the radiation energy density in the plasma, $\rho_{\rm rad}$, 
or via the stress-energy tensor's trace and $\rho_{\rm rad}$. These conceptualisations converge in the formula \cite{Barducci:2020axp}:
\begin{equation}\label{eq:alpha}
	\alpha = \frac{1}{\rho_{\rm rad}} \left.\left( \Delta V - n \,  \frac{\partial \Delta V}{\partial \ln T}\right)\right|_{T = T_n},
\end{equation}
where $\Delta V = V_{\text{eff}}(\langle\phi\rangle_{\text{false}}, T) - V_{\text{eff}}(\langle\phi\rangle_{\text{true}}, T)$ represents the free energy density difference between the two phases, and $n$ assumes values of 1 or $1/4$, corresponding to the definitions based on latent heat and trace anomaly, respectively.}
The inverse duration of the phase transition is approximated as
$\beta = d\ln\gamma/dt$ evaluated at the nucleation time $t_{\text{nuc}}$.
It is convenient to introduce the dimensionless ratio~\cite{Barducci:2020axp}
\begin{equation}\label{eq:beta}
\frac{\beta}{H} = \left.-  \,\frac{d \ln \gamma}{d \ln T} \right|_{T=T_{\rm nuc}}
\simeq -4+ \frac{d}{d \ln T}\left(
\frac{S_{3}}{T} - \frac32 \ln \frac{S_3}{T}\right)_{T=T_{\rm nuc}}
\end{equation}
so that the phase transition happens in $H/\beta$ Hubble times.
The duration of the phase transition gets longer, up to a fraction of Hubble time,
in the regime where $T_{\rm nuc}$ is much smaller than the mass scales of the problem.
This happens because $d(S_3/T)/d\ln T \propto \ln^{-2}(w/T)$ decreases at $T \ll w$, where
the approximation of eq.\eq{VTapprox} gets more accurate.\footnote{\nnn{We do not need to expand
the decay rate density up to higher second order in Taylor series in the time difference from the nucleation time $t_n$,
\begin{equation}
	\label{eq:GammaGen}
    \gamma(t) = \gamma(t_{\rm nuc})\text{exp}[\beta(t-t_{\rm nuc})+ \beta_2(t-t_{\rm nuc})^2+\cdots],
\end{equation}
because, in the considered phase transitions during super-cooling,
the second order coefficient $\beta_2$ negligibly arises by expanding the logarithmic dependence
of $S_3/T$ on $T/w \approx e^{-H(t-t_{\rm nuc})}  T_{\rm nuc} /w \gg 1$.}}
As discussed later in section~\ref{sec:PBHformation}, a substantial amount of primordial black holes forms 
in this regime, near to the critical boundary where nucleation becomes too slow for ending super-cooling.
If $T$ gets smaller than $\Lambda_{\rm QCD}$, super-cooling is ended by QCD effects~\cite{Witten:1980ez,Iso:2017uuu}.


The probability that a point still is in the false vacuum phase at time $t$ is  $\wp(t)=e^{-I(t)}$, where (see e.g.~\cite{Gouttenoire:2023naa,Lewicki:2023ioy,2307.11639})
\be
I(t)=\frac{4 \pi}{3} \int_{t_{\rm c}}^t d t^{\prime} \gamma(t^{\prime}) a^3(t^{\prime}) r^3(t, t^{\prime})
=\frac{4 \pi}{3} \int_T^{T_{\rm c}} \frac{\mathrm{d} T^{\prime} \gamma\left(T^{\prime}\right)}{T^{\prime 4} H\left(T^{\prime}\right)}\left(\int_T^{T^{\prime}} \frac{\mathrm{d} \tilde{T}}{H(\tilde{T})}\right)^3.
\ee
where nucleation starts at $t_{\rm c} \lesssim t_{\rm eq}$ and
$r (t'',t')=\int_{t'}^{t''} dt\, v_{\rm wall}/a$ is the radius at time $t''$ of a bubble formed at time $t'$. 
The percolation temperature $T_{\rm perc}$ is approximatively defined as
$
I(T_{\rm perc}) = 0.34
$
meaning that $34\%$ of the comoving volume has been converted into the true minimum~\cite{Ellis:2018mja,Ellis:2020nnr,2212.07559}. 
While in most of the parameter space $T_{\rm perc}\approx T_{\rm nuc}$, Primordial Black Hole formation will occur
when the phase transition has a long duration $\beta/H \lesssim 8$, such that $T_{\rm perc}\lesssim T_{\rm nuc}$.
The phase transition completes provided that
the  comoving volume ${\cal V}_{\rm false} \propto a(t)^3 \wp(t) $ remaining in the false vacuum decreases~\cite{Ellis:2018mja,Ellis:2020nnr} 
\be\label{eq:crit1}
\left. \frac{d{\cal V}_{\rm false}}{dt}\right|_{t\sim t_{\rm perc}} < 0\qquad\hbox{i.e.}\qquad
\left.3+ \frac{d I(T)}{d \ln T}\right|_{T\sim T_{\rm perc}}<0.
\ee


\subsection{Gravitational waves from the phase transition}
The gravitational wave spectrum generated from first-order phase transitions is dominated by three contributions: 
the collision of true vacuum bubbles, $\Omega_{\text{col}}\,h^{2}$; the propagation of sound waves within the plasma, $\Omega_{\text{sw}}\,h^{2}$; and magnetohydrodynamic turbulence effects, $\Omega_{\text{turb}}\,h^{2}$, as discussed in~\cite{Caprini:2018mtu}. 
Bubble collisions dominate in our scenario, as supercooling leads to a large energy release $\alpha \gg 1$,
The resulting  $\Omega_{\text{col}}\,h^{2}$ is approximated as~\cite{Caprini:2019egz,Helmboldt:2019pan,Barducci:2020axp,Croon:2018new,Weir:2017wfa,Huber:2008hg,Jinno:2016vai}
\be\label{eq:GW_spectr}
\Omega_\text{col}\, h^2 = 1.67 ~ 10^{-5}\,\left(\frac{\beta}{H}\right)^{-2}\, 
\left(\frac{\kappa_\text{col}\,
	\alpha}{1 + \alpha}\right)^{2}\,
\left(\frac{100}{g_{\star}}\right)^{1/3}\,
\Delta_\text{col}(v_{\text{wall} })\,S_\text{col}(f).
\ee
The gravitational wave energy density $\Omega_\text{col}$ achieves its maximal $\alpha$-independent value at $\alpha \gg 1$. 
The coefficient $\kappa_{\text{col}}$ is the efficiency associated to collisions of scalar shells (see e.g.~\cite{Ellis:2018mja}). 
In view of $\alpha\gg1$, we assume $\kappa_{\text{col}} \approx 0.95$~\cite{Lewicki:2019gmv}.
The function of the wall velocity $v_{\rm wall}$ in eq.~(\ref{eq:GW_spectr}) is \cite{Ellis:2020nnr}
 \be\label{eq:GW_delta}
 \Delta_{\text{col}}(v_{\text{wall}})  = 
 \frac{0.48\,v_{\text{wall}}^{3}}{1+ 5.3 \,v_{\text{wall}}^{2}+ 5\, v_{\text{wall}}^{4}}.
 \ee
meaning that larger $v_{\rm wall}$ leads to stronger GW signals. 
In view of $\alpha\gg1$ we can assume highly relativistic bubbles, $v_{{\rm wall}}\simeq 1$ \cite{Ellis:2020nnr,Lewicki:2019gmv,2106.09706}.
The spectral shape $S_i(f)$ in eq.~(\ref{eq:GW_spectr}) is \cite{Croon:2018new,Weir:2017wfa,Huber:2008hg,Jinno:2016vai}
 \begin{eqnarray}\label{eq:GW_shape}
 	S_{\text{col}}(f)  &=& 
 	\left[c_l\left(\frac{f}{f_{\text{col}}}\right)^{-3}\, + (1-c_l -c_h)\left(\frac{f}{f_{\text{col}}}\right)^{-1}\,+ c_h\left(\frac{f}{f_{\text{col}}}\right)\,
 	\right]^{-1}
 \end{eqnarray}
 where $c_l=0.064$ and $c_h=0.48$. 
 $S_{\rm col}$ equals unity at the peak frequency $f_{\text{col}}$ given by 
 \be\label{eq:fpeak}
 f_{\text{col}} =   1.65 \times 10^{-5}\,{\rm Hz}~ \,\Xi_{\text{col}}(v_{\text{wall}})\, \frac{\beta}{H}\, \frac{T_{\rm RH}}{100\,{\rm GeV}} \, \left( \frac{g_\star}{100} \right)^{1/6}\qquad\hbox{for}\qquad T_{\rm RH}\gg T_{\rm nuc}
 \ee
 with~\cite{Caprini:2019egz,Helmboldt:2019pan,Barducci:2020axp} 
 \begin{eqnarray}
 	\Xi_{\text{col}}(v_{\text{wall}})  &=& 
 	\frac{0.35}{1 + 0.069\, v_{\text{wall}} + 0.69\, v_{\text{wall}}^4}.
 \end{eqnarray}

\medskip

Big Bang Nucleosynthesis (BBN) and Cosmic Microwave background (CMB) data provide a bound on the energy density of extra radiation 
(in this case, of gravitational waves).
Such bound is usually presented in terms of an effective number of extra neutrinos~\cite{Maggiore:1999vm}
\beq \Delta N_{\rm eff}=4.4\left.\frac{\rho_{\rm GW}}{\rho_\gamma}\right|_0 =1.8 ~ 10^5
\int_{f_\text{min}}^{\infty} \frac{\text{d}f}{f}   \Omega_\text{GW}(f) h^2 
\approx  10^5     \Omega_\text{GW}^\text{peak} h^2 \label{eq:darkrad2} \eeq
where we can approximate $f_{\rm min}\approx 0$.
The current bound $\Delta N_{\rm eff}\lesssim  0.28$~\cite{Planck:2018vyg}
could be improved by 1 or 2 orders of magnitude with future 
CMB experiments such as~\cite{CMB-S4:2020lpa, CMB-S4:2022ght,Sehgal:2019ewc,CMB-HD:2022bsz}.

\subsection{PBH formation from strong first-order phase transition}\label{sec:PBHformation}
A first order phase transition can lead to formation of black holes~\cite{Liu:2021svg,Kawana:2022olo,Gouttenoire:2023naa,Lewicki:2023ioy} in different ways.
In particular, due to the stochastic nature of nucleation, 
there is a probability that some regions remain for a longer time in the false vacuum 
while the space around them gets filled by true vacuum. 
Re-heating in the true vacuum leads to radiation, so that its energy density $\rho_{\rm true}$ dilutes, 
while the energy density in the false vacuum is $\rho_{\text{false}} \simeq V_\Lambda$.
Thus regions in the false vacuum become relatively denser as quantified by $\delta \equiv \rho_{\text{false}} / \rho_{\text{true}}-1$.
\nnn{Following~\cite{Gouttenoire:2023naa,Gouttenoire:2023pxh} we assume that
a primordial black hole forms when}
the density contrast exceeds the critical value  $\delta_{\text{c}} \approx 0.45$ \cite{2002.12778} in
a region in the false vacuum (approximated as roughly-spherical) with radius
larger than roughly the Hubble parameter.
\nnn{For small perturbations, this value for $\delta_c$ indeed coincides with the one at the horizon-crossing time, and
it does not differ much from that even for the larger perturbations of interest for PBH formation~\cite{1201.2379}. 
It may however mildly vary depending on the perturbation profile~\cite{2404.06547}.
We ignore this issue, since the power spectrum induced by phase transitions does not have sharp features.}

According to~\cite{Gouttenoire:2023naa} the resulting probability that a Hubble patch forms a PBH in
a supercooled first-order phase transition, with $\alpha \gtrsim 100$ and inverse duration  $\beta/H\lesssim 10$, is analytically approximated by 
\begin{eqnarray}\label{eq:Pcol_num}
	\mathcal{P}_{\text {coll }} \approx \exp\bigg[-a\bigg(\frac{\beta}{H}\bigg)^{b}(1 + \delta_{\text{c}})^{c\,{\beta}/{H}}\bigg].
\end{eqnarray}
where $a =  0.5646$, $b = 1.266$ and $c= 0.6639$. 
The collapse probability does not depend on $\alpha$ in the limit of strong supercooling $\alpha \gg 1$ 
where the residual initial radiation energy is negligible.
The collapse probability crucially depends on the duration of the phase transition $\beta/H$,
as a small enough $\beta/H$ 
(i.e. a long enough duration) allows Hubble-sized late-blooming regions.
The resulting fraction of the density of PBH with respect to the density of dark matter is
\begin{eqnarray}
f_{\text{PBH}} \equiv\frac{ \rho_{\text{PBH}}}{\rho_{\text{DM}}} \approx \frac{	\mathcal{P}_{\text {coll }} }{2.2 \times 10^{-11}} \frac{T_{\text{eq}}}{140\, \text{GeV}}.
\end{eqnarray}
where $\mathcal{P}_{\text {coll }}$ is given in eq.~(\ref{eq:Pcol_num}), and $T_{\rm eq}$ is the temperature at which vacuum and radiation have the same density, eq.\eq{Teq}.
So $f_{\rm PBH}\sim 1$ roughly corresponds to $\beta/H \sim 8$.
This value is realized when the couplings 
of the model are mildly small and the tunnelling is near to being too slow for the phase transition to complete,
such that the nucleation temperature is low.

PBH can explain all DM, $f_{\rm PPH}=1$, in the mass range (see~\cite{Cirelli:2024ssz} for a review)
\beq 10^{-16}\,  M_\odot \lesssim M_{\text{PBH}} \lesssim 3\times10^{-12}\, M_\odot .
\label{eq:PBHmassrange}\eeq 
PBH form with mass roughly given by the mass within the sound horizon volume at bubble collision time,
\be
M_{\text{PBH}} \approx  F \frac{4\pi}{3} \left(\frac{c_s}{H}\right)^3 \rho_{\rm rad}^{\rm late}\bigg|_{t_{\rm coll}} \approx
M_{\odot} \left(\frac{ 20 
}{g_{\star}(T_{\text{eq}})}\right)^{1/2} \left(\frac{0.14\, \text{GeV}
}{T_{\text{eq}}}\right)^{2}
\ee
where $g_{\star} \approx 100$, $M_{\odot} \approx 2\times 10^{30}\,\text{kg}$ is the solar mass, 
$c_s\simeq 1/\sqrt{3}$ is the sound speed, and \nnn{$F\lesssim 1$ is a numerical factor which depends on the details of the gravitational collapse~\cite{2103.03867}}.
$\rho_{\rm rad}^{\rm late}$ is the false vacuum energy in a late-patch~\cite{Gouttenoire:2023naa}.
The DM range in eq.\eq{PBHmassrange} roughly corresponds to equilibrium temperatures $50 \TeV \lesssim T_{\rm eq} \lesssim 10\,{\rm PeV}$.

\subsubsection*{Spin of the primordial black holes}
It is interesting to assess the small initial spin of the PBH since mechanisms for its growth have been explored~\cite{Jaraba:2021ces,Hofmann:2016yih,Calza:2021czr}.
The spin of the PBH is parameterized by the dimensionless Kerr parameter $a_*$, with variance approximated by~\cite{Yoo:2018kvb,DeLuca:2019buf,Harada:2020pzb,2311.03406,Conaci:2024tlc}
\bea
\langle a_*^2 \rangle^{1/2} \simeq 
\frac{4.0 ~ 10^{-3}  \sqrt{1-\gamma^2}}{1 + 0.036 \left[ 21 - 2 \log_{10} \left( {f_{\rm PBH}}/{10^{-7}} \right) - \log_{10} \left( {M_{\rm PBH}}/{10^{15} {\rm g}} \right) \right]} \, .
\label{eq:spin}
\eea
The parameter $\gamma$ typically lies in the range $0.81-1$; we assume a reference value $\gamma = 0.96$~\cite{DeLuca:2019buf,Escriva:2022duf,Harada:2020pzb}.
\nnn{In general, $\gamma$ is inversely proportional to the spectral width $\sigma$ of power spectrum \cite{2311.03406,2409.06494},
taking into account all scales. 
In our case, scales smaller than the horizon do not contribute to the creation of PBHs. 
This leads to a monotonically increasing functional dependence of the PBH abundance on $\sigma$. 
The assumed value of $\gamma$ should be realistic in the physically relevant domain, where the PBH abundance constraints are satisfied~\cite{2311.03406,2409.06494}.
In our scenario, the scalar avoids a phase where its quanta dominate the energy density:
it quickly transfers its energy to the SM radiation, opening the phase when PBHs are formed.
}

\section{Minimal model with singlet scalar DM}\label{sec:mods'}
We now start to compute the signals of various models.
We start from a minimal model of dynamical electroweak symmetry breaking,
that involves two extra scalars $s$ and $s'$ besides the Standard Model Higgs doublet $H$. 
Assuming, for simplicity, that the  theory is separately invariant under $s\to - s$ and $s'\to - s'$, the discrete symmetry  $\mathbb{Z}_2\otimes\mathbb{Z}'_2$
allows the scale invariant potential
\beq\label{eq.Vtree}
V_{\rm tree} =V_\Lambda +\lambda_H|H|^4+\frac{\lambda_S}{4} s^4+\frac{\lambda_{S^{\prime}}}{4} s^{\prime 4}+\frac{\lambda_{H S}}{2}|H|^2 s^2  +\frac{\lambda_{H S^{\prime}}}{2}|H|^2 s^{\prime 2}+\frac{\lambda_{S S^{\prime}}}{4} s^2 s^{\prime 2} .
\eeq
A constant term $V_\Lambda$ is added such that $V=0$ at its minimum. Effects due to the explicit breaking of the scale symmetry are Planck-suppressed.
The scalar $s'$ has been introduced such that the couplings $\lambda_S$ and $\lambda_{HS}$ run with $\beta$ functions
\beq  \beta_{\lambda_S} \simeq \frac{ \lambda_{SS'}^2}{2(4\pi)^2},\qquad
\beta_{\lambda_{HS}}\simeq \frac{\lambda_{SS'} \lambda_{HS'}}{(4\pi)^2}\eeq
that can make them negative at lower energy,
inducing dynamical symmetry breaking.
This theory admits multiple symmetry-breaking patterns depending on which combination of couplings crosses zero first.
Following~\cite{2204.01744} we define a parameter $R$ that controls which effective running coupling turns negative first:
\beq
\lambda_S^{\rm eff}(s) = \frac{{\beta}_{\lambda_S }}{2}   \ln \frac{s^2}{e^{1/2}w^2 },\qquad
\lambda_{HS}^{\rm eff}(s) = \frac{{\beta}_{\lambda_{HS} }}{2} \ln \frac{R s^2 }{w^2}.
\eeq
We are interested in the phase where $s$ and $h$ acquire vacuum expectation values,
$\langle s\rangle = w \gg \langle h \rangle = v=246.2$ GeV.
This phase happens for $R\lesssim  1$ when 
couplings cross the critical boundary 
$\lambda_{H S}=-2 \sqrt{\lambda_H \lambda_S}$ while  $\lambda_{HS}$ is negative.
The standard Gildener- Weinberg approximation applies at $R \ll 1$, such that $\lambda_{HS}$ has a roughly constant value. 
Phenomenology remains viable when $R$ is mildly smaller than 1 and multi-phase effects (encoded in the running of $\lambda_{HS}$)
become relevant~\cite{2102.01084,2204.01744}.


\begin{table}[t]
\begin{center}
\begin{tabular}{l|cccc}
            {{Benchmark points}}& \textbf{\color{blue}A} & \textbf{\color{red}B} & \textbf{\color{orange}C} & \textbf{\color{green}D} \\
\hline        \hline
Scalon mass $m_{s}$ in TeV & $59.5$ & $200$ &  $10^{3}$ & $ 10^{4}$ \\
DM mass $m_{s'}$ in PeV & $0.84$ & $2.72$ & $12.7$ & $114$ \\
            \hline
Equilibrium temperature $T_{\rm eq}$ in TeV & $59$ & $190$ & $890$ & $7996$ \\
Nucleation temperature $T_{\rm nuc}$ in GeV & $0.3$ & $2.7$ & $47$ & $2600$ \\
Percolation temperature $T_{\rm perc}$ in GeV & $0.1$ & $1.1$ & $29$ & $2580$ \\
Duration of the phase transition $\beta/H_{n}$ & $6.5$ & $6.6$ & $6.7$ & $6.9$ \\
 \hline
 PBH mass $M_{\text{PBH}}/M_{\odot}$ & $2.5 ~ 10^{-12}$ & $2.4 ~ 10^{-13}$  & $1.1 ~ 10^{-14}$ & $1.4 ~ 10^{-16}$ \\
 PBH abundance            $f_{\text{PBH}} = {\rho_{\rm PBH}}/{\rho_{\rm DM}}$ & 
 $\approx1$ & $ \approx1$ &  $ \approx1$ & $ \approx1$ \\
 Tuning $\Delta = d \ln f_{\rm PBH}/d\ln m_{s,(s')} $ & 935 (970) & 966 (1004) & 1008 (1052) &  1061 (1115)  \\  
PBH spin $\left\langle a_*^2\right\rangle^{1 / 2}$ & $0.00111$ & $0.00107$ & $0.00102$ & $0.00096$ \\
\hline
GW abundance $\Omega_{\text{col}}^{\text{peak}} h^2$ & $1.31~ 10^{-8}$ & $1.29 ~ 10^{-8}$ & $1.24~ 10^{-8}$ & $1.19~ 10^{-8}$ \\
GW frequency $f_{\text{col}}^{\text{peak}}$ in Hz & $7 ~ 10^{-8}$ & $6.3 ~ 10^{-7}$ & $1.1~ 10^{-5}$ & $6.3~ 10^{-4}$ \\
 \end{tabular}
\caption{\label{tab:BP}\em Benchmark points. 
The first two rows, together with $\ln R = -10$, define the parameter values.
The other rows show the resulting predictions.
Fig.~\ref{fig:plotGW} illustrates the associated gravitational waves, and
fig.~\ref{fig:plotPBH} illustrates the associated PBH mass and abundance. }
\end{center}
\end{table}

As the Higgs mass is known to be $m_h\approx 125.1\,{\rm GeV}$, we use as free parameters the masses of $s$ and $s'$ and $R$.
The various couplings are then approximated in terms of our parameters as~\cite{2204.01744}
\beq
  \lambda_{SS'} &\approx \frac{(4 \pi)^{2} m_{s}^{2}}{m_{s'}^{2}},\qquad
  \lambda_{HS'} &\approx -\frac{(4 \pi)^{2} m_{h}^{2}}{m_{s'}^{2} \ln R},\qquad
  w\simeq \frac{\sqrt{2} m^2_{s'}}{4\pi m_s}.
\eeq
Furthermore, $s'$ is a stable DM candidate
in the phase with $\langle s'\rangle=0$.
The $s'$ relic abundance can match the DM abundance in two different ways:
at moderately large coupling via freeze-out at $T_{\rm dec} \approx m_{s'}/25$;
at small coupling via super-cooling~\cite{1805.01473}.
We focus on the first regime.
The re-heating temperature $T_{\rm RH}$ of eq.\eq{TRH}
is $T_{\rm RH}\approx 0.07 m_{s'}$,
a factor of 2 larger than $T_{\rm dec}$.
So this freeze-out regime is realized.
The $s$-wave DM annihilation cross section $\sigma_0$ and the spin-independent direct detection cross section are~\cite{2204.01744}
\beq
\sigma_0 \simeq \frac{ \lambda_{SS'}^2+4\lambda_{HS'}^2}{64\pi m_{s'}^2},\qquad
\sigma_{\rm SI}\simeq \frac{64 \pi^3 f_N^2 m_N^4}{m_{s'}^6}
\eeq
with $f_N \approx 0.3$ and $m_N \approx0.946\,{\rm GeV}$.
The cosmological DM abundance is reproduced when $\sigma_0 \approx 1/M^2$ with $M \approx 23\TeV$.
Current bounds on direct detection are satisfied for $m_{s'}\gtrsim2\,{\rm TeV}$~\cite{2204.01744,LZlast}.
\nnn{Then indirect detection bounds are satisfied}~\cite{Cirelli:2024ssz}.

As PBH provide a second DM candidate, we will also consider the possibility that $s'$ is unstable,
such that $s'$ no longer is DM, and acquires more visible collider signals.
Destabilization of $s'$ happens if the $\mathbb{Z}_2$ symmetry is broken either spontaneously by $\lambda_{SS'}<0$,
or explicitly.  For example,  $s$ or $s'$ might linearly couple to
right-handed neutrinos allowing to generate neutrino masses~\cite{1807.11490}.

\begin{figure}[t]
\begin{center}
\includegraphics[width=.6\textwidth]{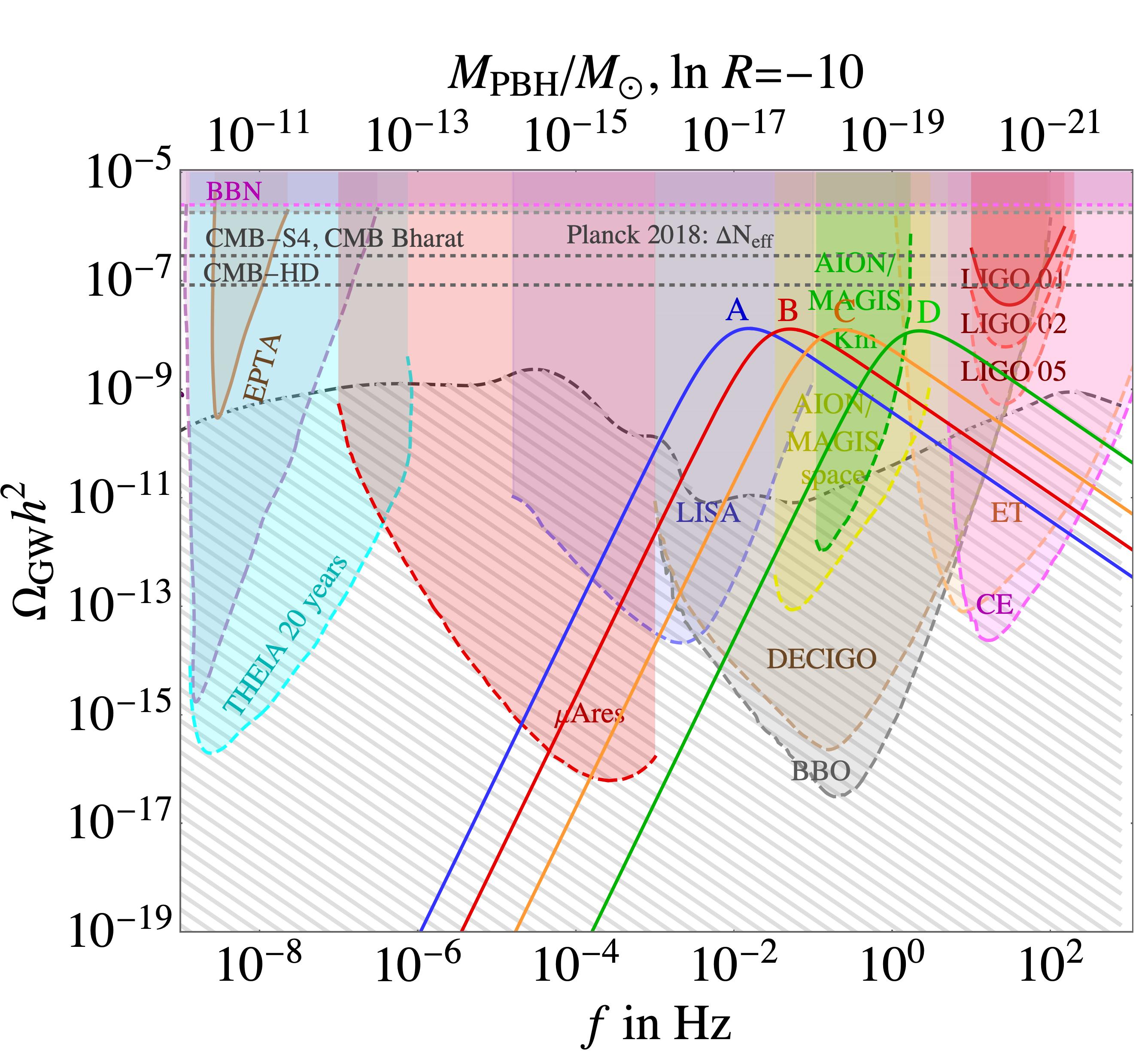}  
\caption{\label{fig:plotGW}\em Gravitational wave spectrum as a function of frequency $f$ at the benchmark points A, B, C, D listed in table~\ref{tab:BP}. 
The predicted gravitational wave signals (colored curves)
fall within the sensitivity curves of possible future gravitational wave detectors (dashed shaded regions).
The upper axis shows the mass of the corresponding primordial black holes.
The gray hatched curve \nnn{indicates} the expected \nnn{but significantly uncertain} astrophysical foreground.
}
\end{center}
\end{figure}

The thermal potential is approximated by eq.\eq{VTapprox} with
\beq m^2 = \left(\frac{\lambda_{SS'}}{2} +2 \lambda_{HS}\right)\frac{T^2}{12},\qquad 
k =  \frac{\lambda_{SS'}^{3/2} +4\lambda_{HS}^{3/2} }{2\sqrt{2}}\frac{T}{4\pi},\qquad
\lambda =\beta_{\lambda_S}\ln \frac{w}{T}
\eeq 
The phase transition is ended by nucleation if 
$S_3/T \sim 16\sqrt{6}\pi^3 \lambda_{SS'}^{-3/2}/\ln(w/T)>4\ln M_{\rm Pl}/T$,
implying a not too small coupling $\lambda_{SS'}$~\cite{1910.13460}.
While this rough estimate qualitatively explains the main results,
gravitational wave and primordial black holes are more precisely computed numerically.

%
%

\begin{figure}[t]
\begin{center}
\includegraphics[width=.6\textwidth]{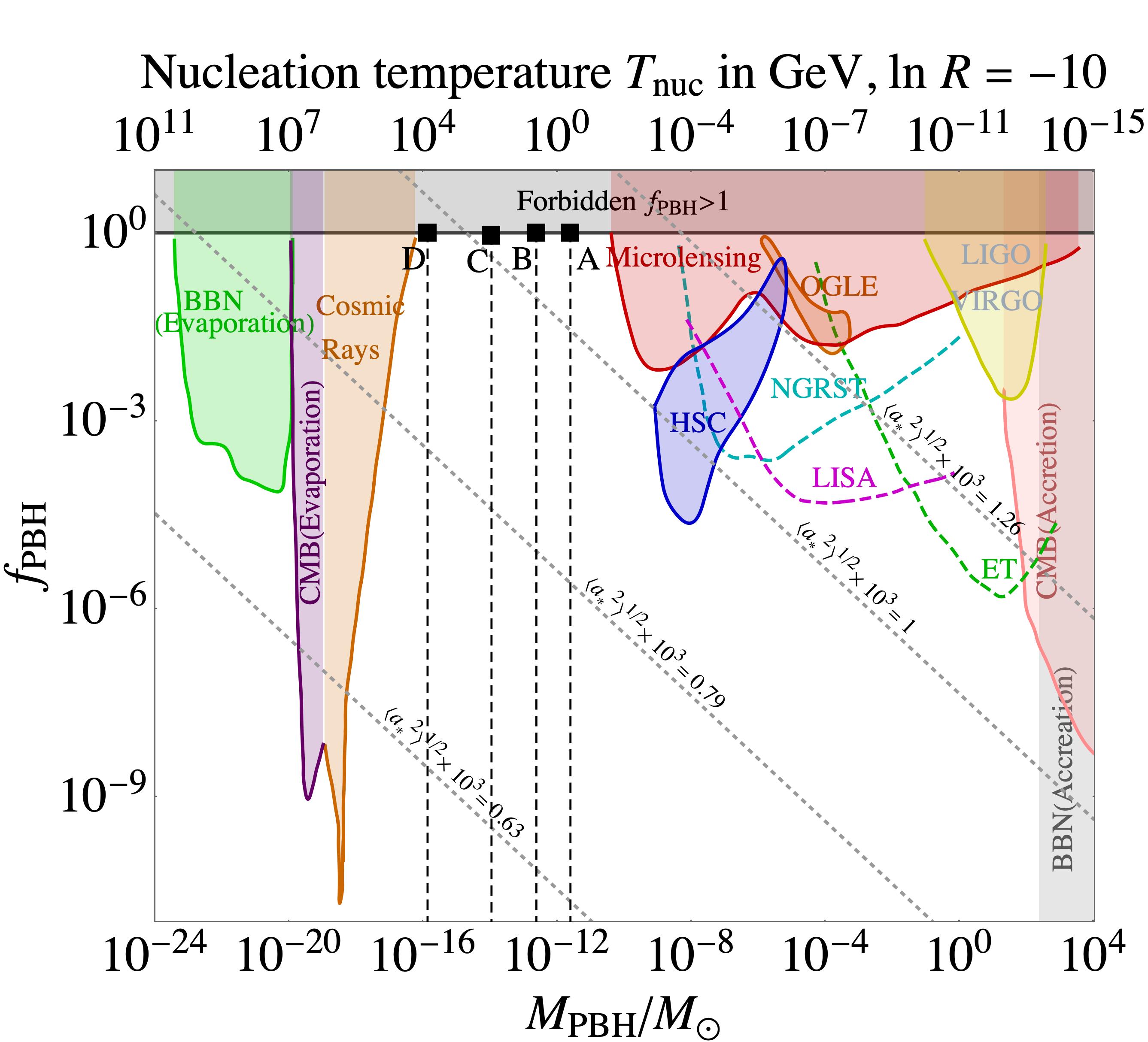} 
\caption{\em\label{fig:plotPBH}
Prediction of the A, B, C, D benchmark points of table~\ref{tab:BP}
for the primordial black holes mass and for their cosmological abundance 
$f_{\rm PBH}$ in units of the DM abundance.
The colored regions with solid borders are excluded.
Dashed curves indicate future experimental sensitivities. 
Diagonal dotted lines represent different values of PBH spin.}
\end{center}
\end{figure}

\subsection{Primordial Black Holes as DM}
We start the analysis by focusing in table~\ref{tab:BP} on some benchmark point in parameter space, all with $\ln R=-10$,
and with $m_s$, $m_{s'}$ selected such that nucleation happens after a significant amount of super-cooling.
Then, all points predict significant comparable amount of GW, $\Omega_{\rm GW}\sim 10^{-8}$.
As shown in fig.\fig{plotGW} the GW frequency spectra mostly differ by position of the peak, given by eq.\eq{fpeak}.
The peak frequency is lower at larger values of $m_{s,s'} \gg v$, which remains compatible with generating the weak scale $v$
thanks to small enough values of $\lambda_{HS}$.
Fig.\fig{plotGW} also shows the expected sensitivity reaches of possible future GW observatories,  broadly classified as: 
\begin{itemize}
    \item {Ground-based  interferometer detectors} at higher $f$:
    a\textsc{LIGO}/a\textsc{VIRGO} (red dashed)~\cite{LIGOScientific:2014pky,VIRGO:2014yos,LIGOScientific:2019lzm}, \textsc{AION}~\cite{Badurina:2019hst} (orange solid), \textsc{Einstein Telescope (ET)}~\cite{Punturo:2010zz,Hild:2010id} (blue solid), \textsc{Cosmic Explorer (CE)} ~\cite{LIGOScientific:2016wof,Reitze:2019iox} (blue dashed). 
LIGO-VIRGO provide the only  existing bound~\cite{KAGRA:2021kbb}. 
We also show the projected sensitivity at the end of the next advanced LIGO-VIRGO phase~\cite{LIGOScientific:2014pky,VIRGO:2014yos,LIGOScientific:2019lzm}. 

\item   {Space-based interferometers}:  \textsc{LISA}~\cite{LISA:2017pwj,Baker:2019nia} (pink solid), \textsc{BBO}~\cite{Crowder:2005nr,Corbin:2005ny} (green dashed),     \textsc{DECIGO}/\textsc{U-DECIGO}\cite{Yagi:2011wg,Kawamura:2020pcg} (green solid), \textsc{AEDGE}~\cite{AEDGE:2019nxb,Badurina:2021rgt} (orange dashed), \textsc{$\mu$-ARES}~\cite{Sesana:2019vho} (magenta dashed).
These probe the mHz frequency range in which PBH from a supercooled FOPT can explain all the DM abundance.
    
 \item Lower frequencies are probed by {recast of astrometry proposals} 
 \textsc{GAIA}/\textsc{THEIA}~\cite{Garcia-Bellido:2021zgu} (brown dashed), and by
 {pulsar timing arrays}: \textsc{SKA}~\cite{Weltman:2018zrl} (purple), \textsc{EPTA}~\cite{Lentati:2015qwp,Babak:2015lua} (purple dashed),   
 \textsc{NANOGrav}~\cite{NANOGrav:2023gor} (blue shaded region).
\end{itemize}
The GW signals are above the expected astrophysical backgrounds, shown by the hatched curve in fig.\fig{plotGW}.
\nnn{This is obtained summing the different expected sources, and is subject to an uncertainty by about an order of magnitude. 
We have not taken into account that future observatories are expected to be able to partially subtract the astrophysical backgrounds.
These issues have been studied in many papers. 
For a short summary and a list of references see e.g.~\cite{2311.16236}.}

\medskip

Next, fig.\fig{plotPBH} illustrates the predicted mass of abundance of Primordial Black Holes at the benchmark points.
We choose points within the mass range where PBH can be all of DM, and fixed the parameters $m_{s,s'}$ such that 
PBH match the cosmological DM abundance, $f_{\rm PBH} = 1$.
In this region the PBH abundance has a strong sensitivity to model parameters,
\beq \Delta \equiv \frac{\partial \ln f_{\rm PBH}}{\partial \ln m_{s,s'} }\sim1000.\eeq
Table~\ref{tab:BP} shows precise sensitivity values at the benchmark points.
A similar sensitivity arises in theories where PBH are produced via enhanced primordial inhomogeneities
(altought specific models can be more tuned).

Fig.\fig{plotPBH} also depicts existing constraints on $f_{\rm PBH}$ (see~\cite{Green:2020jor,Saha:2021pqf,Laha:2019ssq,Ray:2021mxu} for details on constraints):
\begin{itemize}
\item {Hawking evaporation of PBH} is relevant at lower $M_{\rm PBH}$ and  implies constraints using data from CMB\,\cite{Clark:2016nst}, EDGES\,\cite{Mittal:2021egv},  {\sc Integral}\,\cite{Laha:2020ivk,Berteaud:2022tws}, {\sc Voyager}\,\cite{Boudaud:2018hqb}, 511\;keV\,\cite{DeRocco:2019fjq},
EGRB\,\cite{Carr:2009jm}.

\item {Micro-lensing} observations from HSC~\cite{Niikura:2017zjd}, 
EROS\,\cite{EROS-2:2006ryy}, {\sc Icarus}\,\cite{Oguri:2017ock}, including a PBH hint of PBH from OGLE\,\cite{Niikura:2019kqi}.
The dashed curve shows the micro-lensing future sensitivity of the NGRST~\cite{DeRocco:2023gde}.

\item The range  $M_{\rm PBH}\sim M_{\odot}$ is constrained by LIGO-VIRGO-KAGRA (LVK) observations of PBH-PBH mergers~\cite{Franciolini:2022tfm,Kavanagh:2018ggo,Hall:2020daa,Wong:2020yig,Hutsi:2020sol,DeLuca:2021wjr,Franciolini:2021tla}. 
Future GW interferometers like ET and LISA are expected to reach the sensitivities depicted as dashed curves~\cite{DeLuca:2021hde,Pujolas:2021yaw,Franciolini:2022htd,Martinelli:2022elq,Franciolini:2023opt,Branchesi:2023mws}.
PBH are expected to accrete leading to extra constraints adopted from~\cite{Serpico:2020ehh,Piga:2022ysp}. 
\end{itemize}

\subsection{General parameter space of the model}
\begin{figure}[p]
$$\includegraphics[width=0.35\textwidth,height=0.35\textwidth]{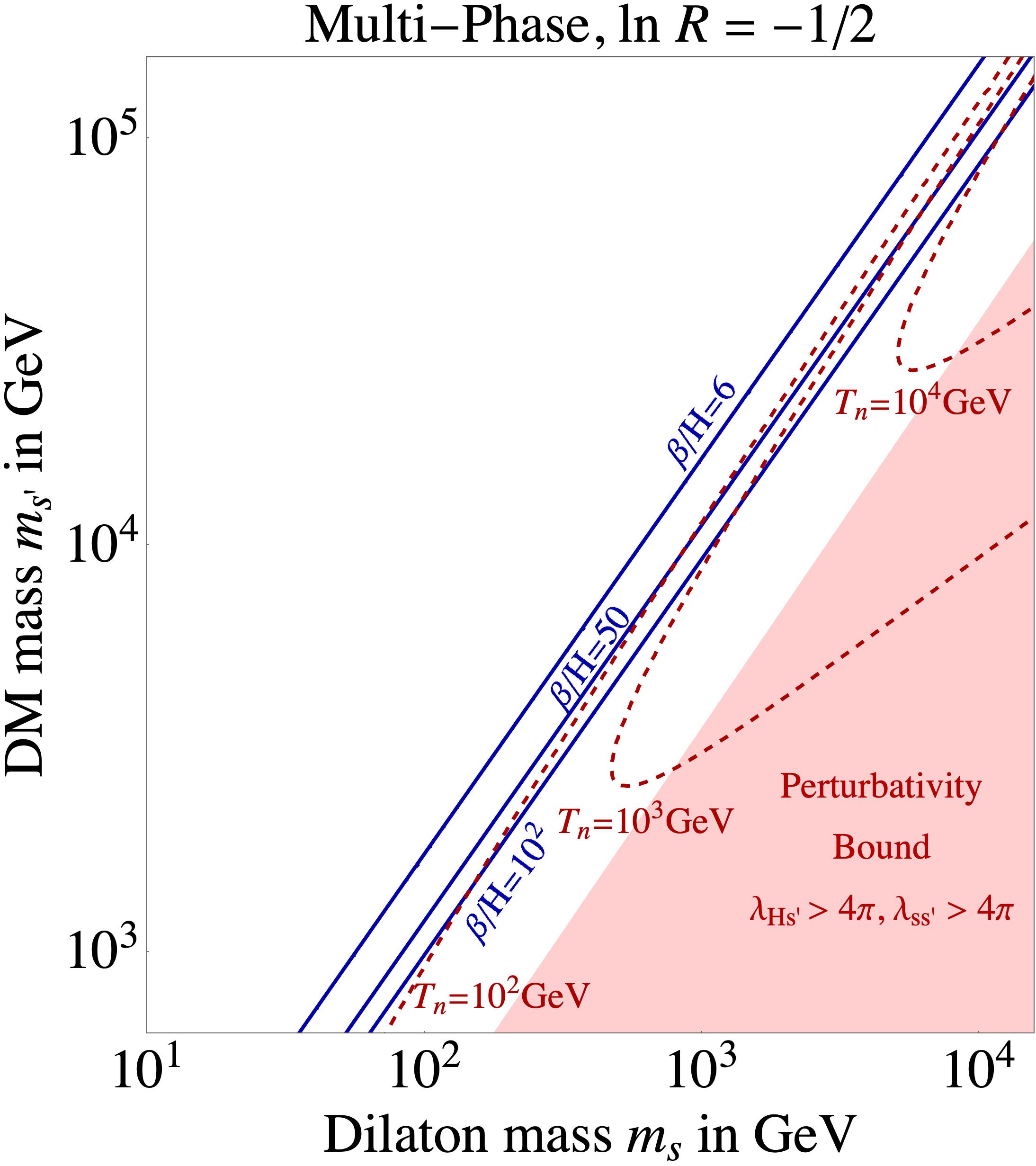}\qquad\qquad
\includegraphics[width=0.35\textwidth,height=0.35\textwidth]{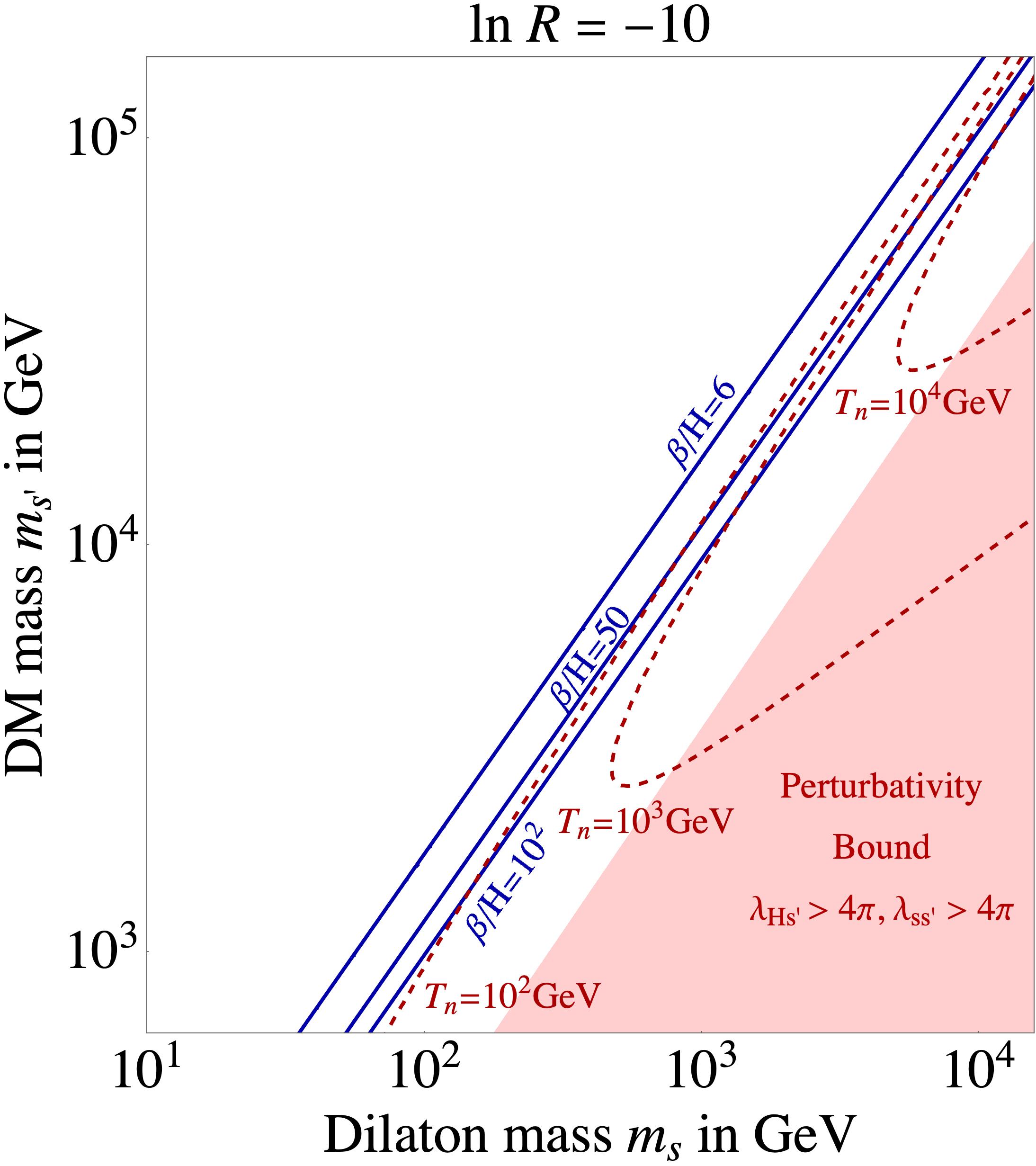}$$
\vspace{-2ex}
\caption{\label{fig:phiend}\em Contour levels of the nucleation temperature $T_{\rm nuc}$ (red dashed curves)
and of the inverse duration of the phase transition $\beta/H$.
A significant amount of PBH are produced at $\beta/H \lesssim 8$.}
$$\includegraphics[width=0.35\textwidth,height=0.35\textwidth]{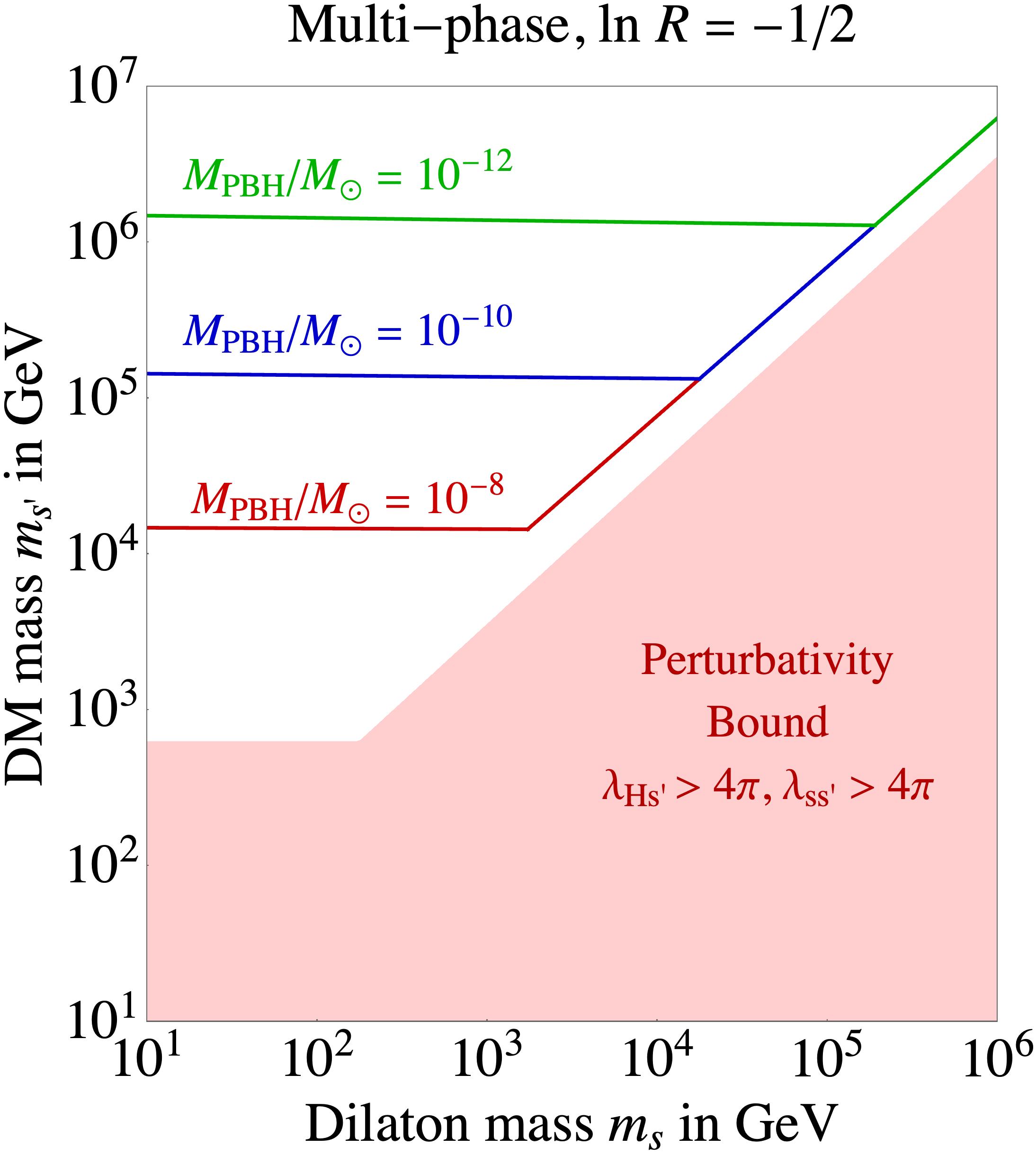}\qquad\qquad
\includegraphics[width=0.35\textwidth,height=0.35\textwidth]{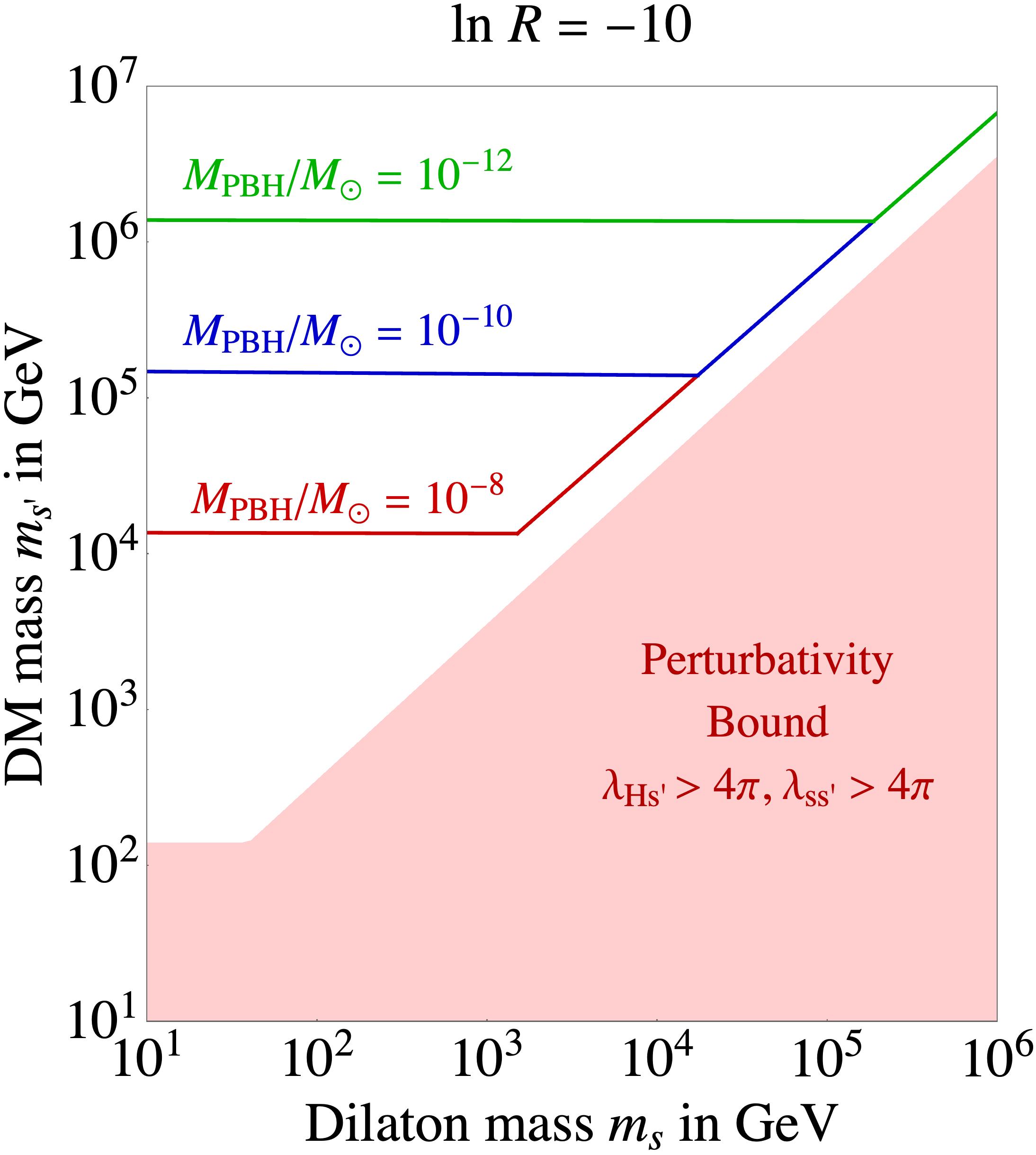}$$
\vspace{-2ex}
\caption{\em \label{fig:MPBHfig} Contour-levels of the mass of produced primordial black holes.  }
$$ \includegraphics[width=0.35\textwidth,height=0.35\textwidth]{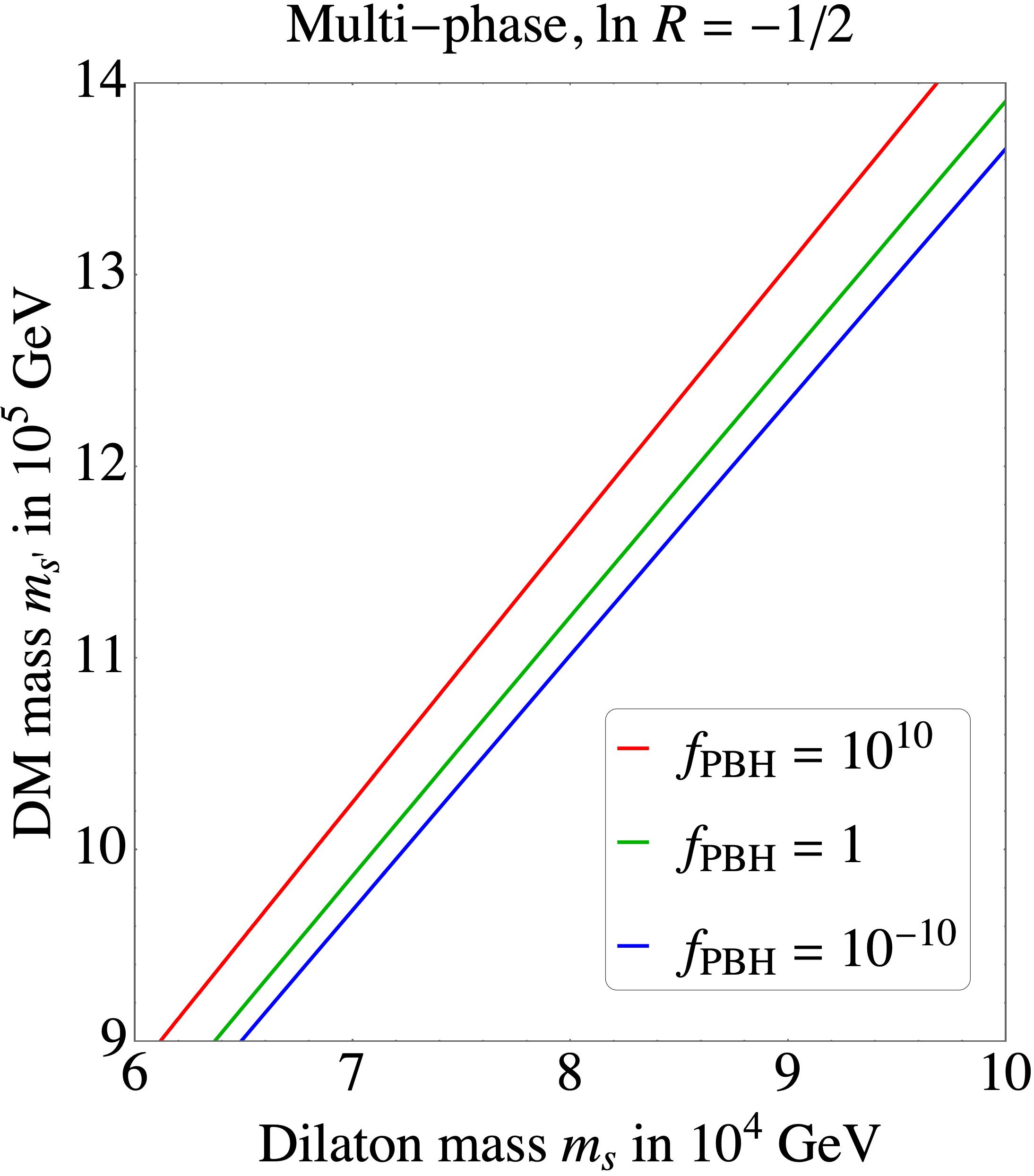}\qquad\qquad
 \includegraphics[width=0.35\textwidth,height=0.35\textwidth]{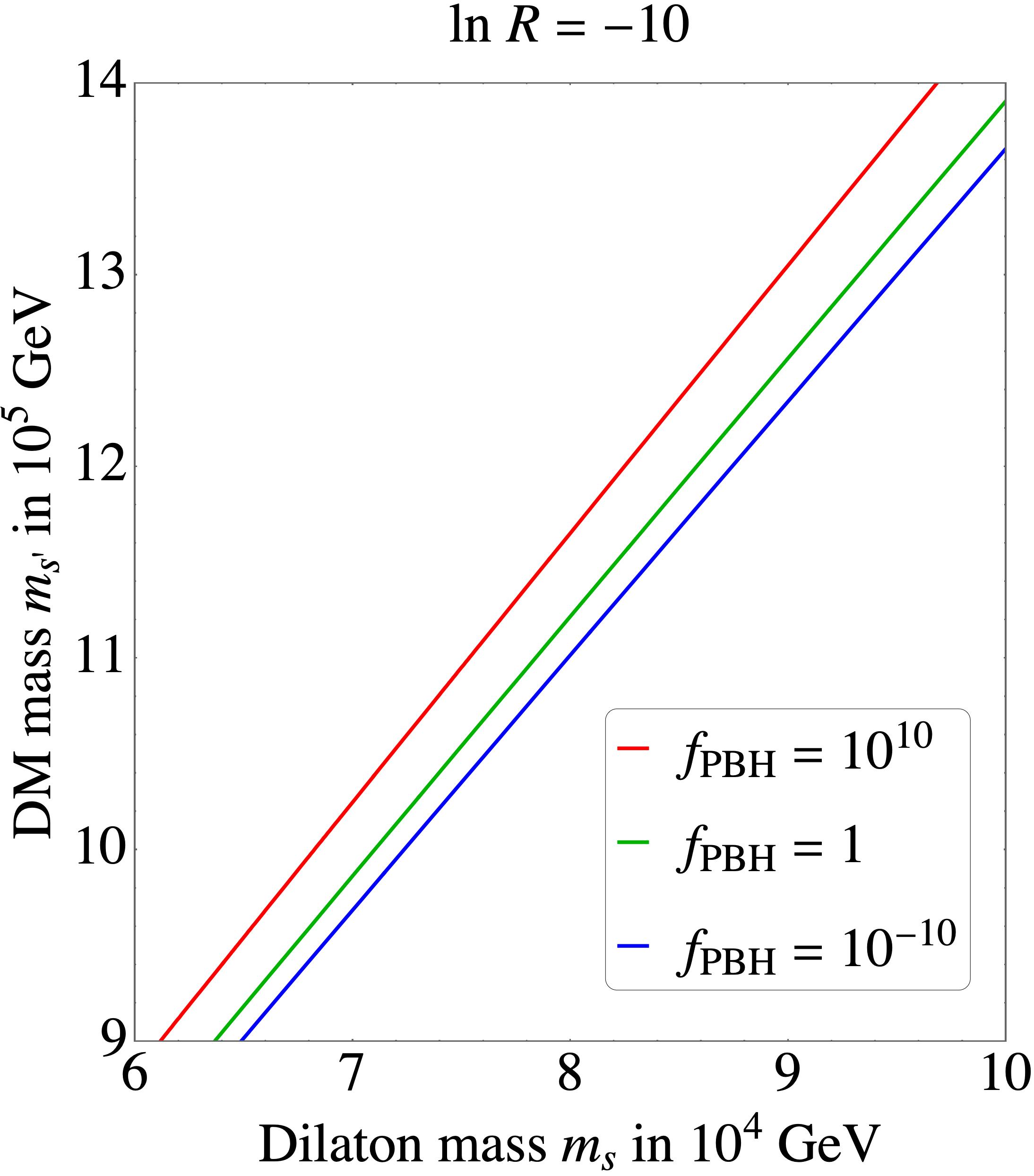}$$
 \vspace{-2ex}
 \caption{\em  \label{fig:fBHfig} Contour levels of the cosmological abundance of Primordial Black Holes.  }
   \end{figure}

We next move from the benchmark points to explore the full parameter space of the model.
We show plots as function of the DM mass $m_{s'}$ and of the dilaton mass $m_{s}$ at fixed values of $\ln R$, focusing on two values:
\begin{itemize}
\item $\ln R=-10$, in  the Gildner-Weinberg regime;
\item $\ln R = - 1/2$, where $\lambda_{HS}$ is as small as allowed by its running and multi-phase effects become relevant~\cite{2204.01744}.
\end{itemize}
Fig.\fig{phiend} shows contour levels of the nucleation temperature and of the inverse duration $\beta/H$
of the phase transition.
Interesting values are obtained in a region ranging from too large non-perturbative couplings down to too small couplings, 
where the phase transition is ended in a too rapid way by QCD rather than by nucleation~\cite{Gouttenoire:2023pxh,2408.03649}.
Fig.\fig{MPBHfig} shows, in the same plane, contour values of the PBH masses.
Fig.\fig{fBHfig} shows contour values of the PBH abundance.
As it is exponentially sensitive to parameter values, the figures focuses on a small portion of the parameter
space where the PBH abundance is around the DM abundance. 
\nnn{In this region, PBH can drive a small-scale clustering of particle DM, thereby enhancing indirect detection signals.}

\smallskip

\begin{figure}[t]
$$\includegraphics[width=.45\textwidth]{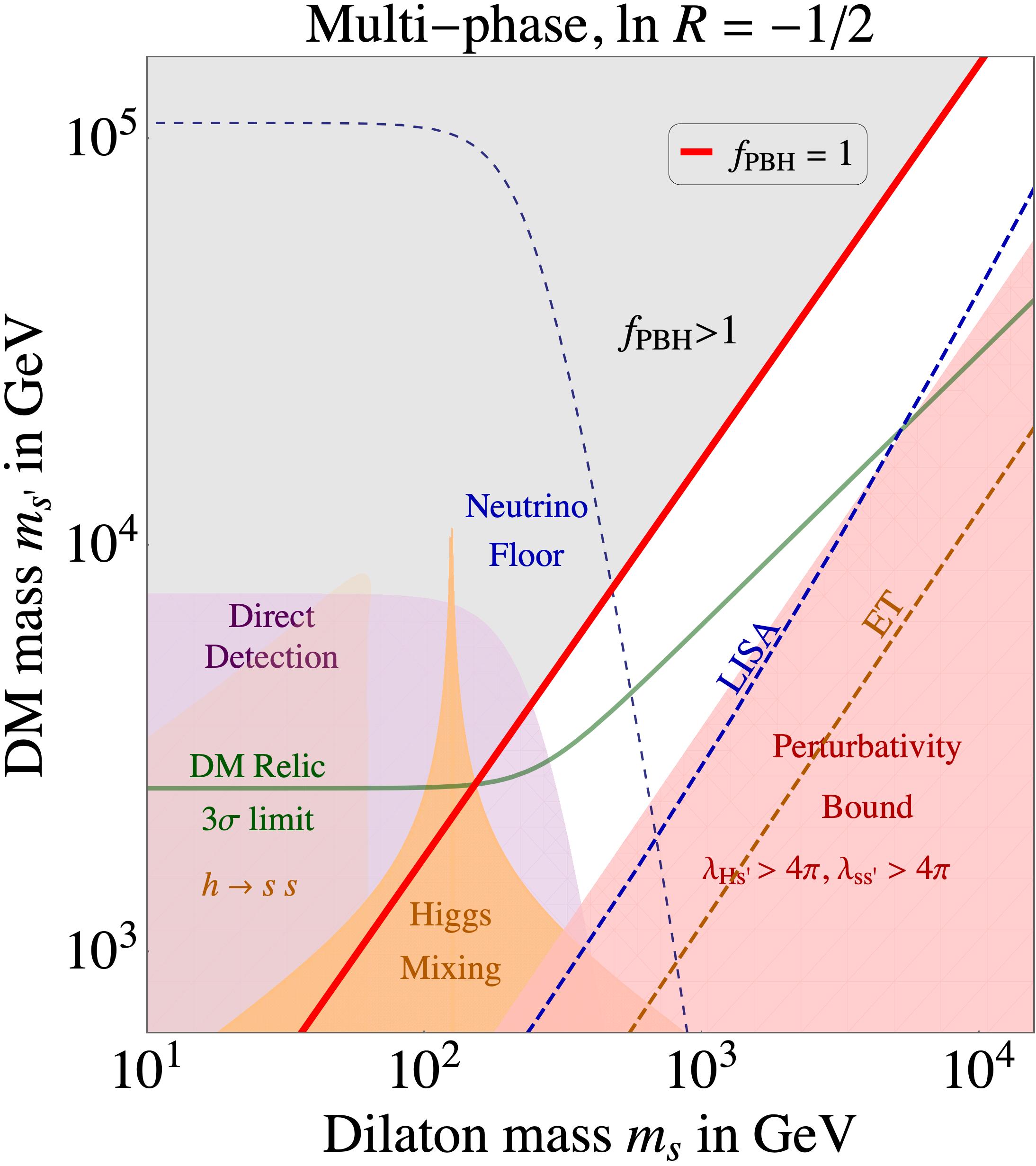}\qquad
    \includegraphics[width=.45\textwidth]{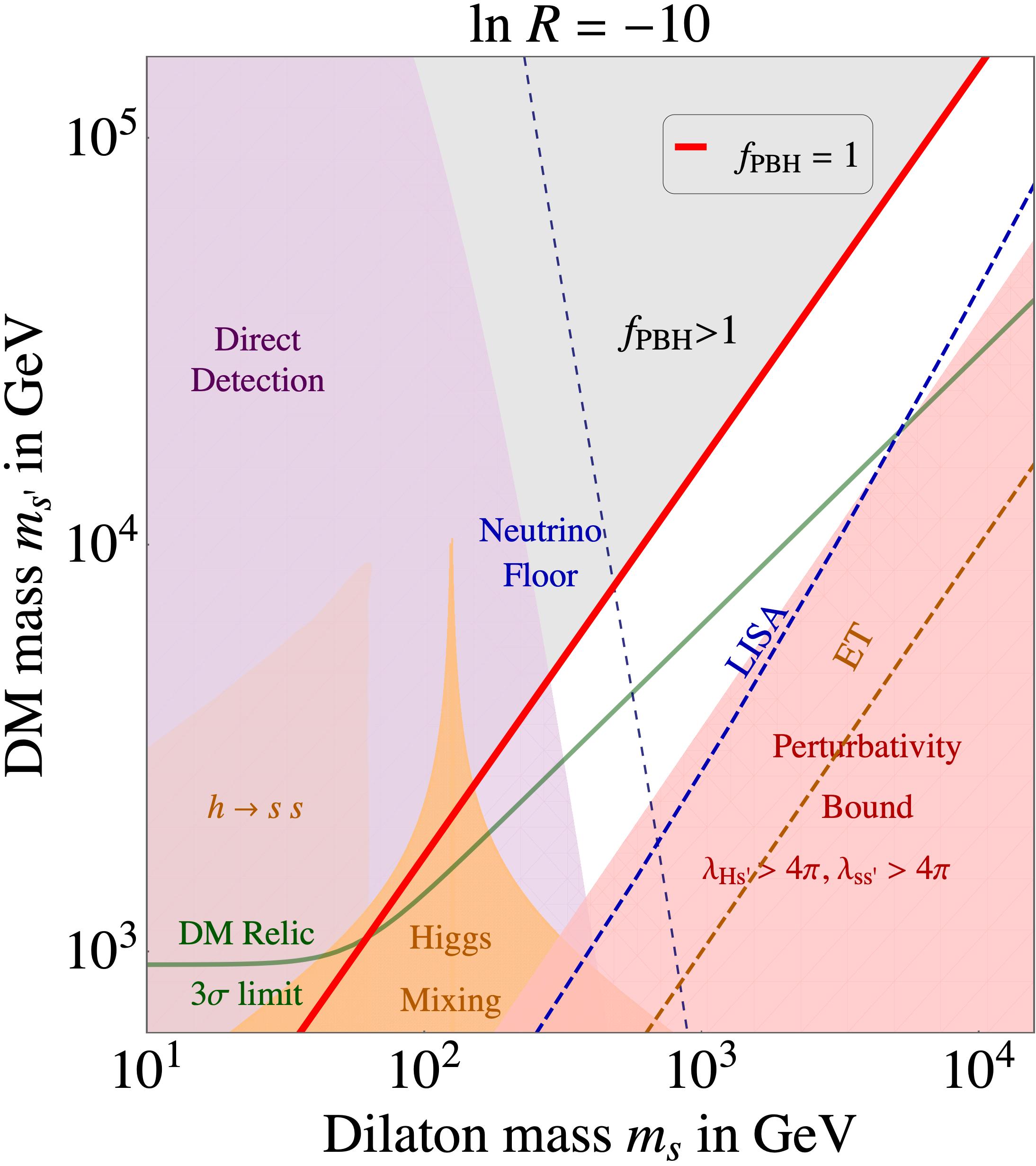}$$
\caption{\label{fig:mainfig}\em Summary plot for the minimal model.
The right panel assumes a  larger mixed quartic $\lambda_{HS}$, 
while in the left panel $\lambda_{HS}$ is so small that its running leads to multi-phase effects. 
The cosmological relic abundance of particle (of Primordial Black Hole) DM is over-abundant above the green (red) curve.
Gravitational waves appear detectable above the curves indicated as LISA and ET.
The shaded regions are excluded by DM direct detection (purple);
Higgs $h\to ss$ decays (beige); 
bounds on Higgs/dilaton mixing (orange);
too large quartic couplings (pink).
}
\end{figure}

Next, we overlay various main results in fig.\fig{mainfig}.

Assuming that $s'$ is stable, we here show its relic freeze-out abundance:
it matches the cosmological DM abundance along the green curve, and gets larger above the green curve.
Furthermore, the PBH abundance matches the DM abundance along the red curve, and gets larger above.
In general, the abundances of the two DM candidates (PBH and $s'$) should be summed.
The plot shows that $s'$ particle DM dominates over PBH in most of the allowed parameter range
(unshaded region in fig.\fig{mainfig}).
The portion around $m_s\sim 200\GeV$ where $f_{\rm PBH}\sim 1$ has $M_{\rm PBH} \sim 10^{-5}M_\odot$, 
above the mass range where PBH can be all DM.
Assuming instead that $s'$ is unstable, PBH provide DM in a different portion of the parameter space, along the red curve.

\begin{figure}[t]
$$    
  \includegraphics[height=7cm,width=7cm]{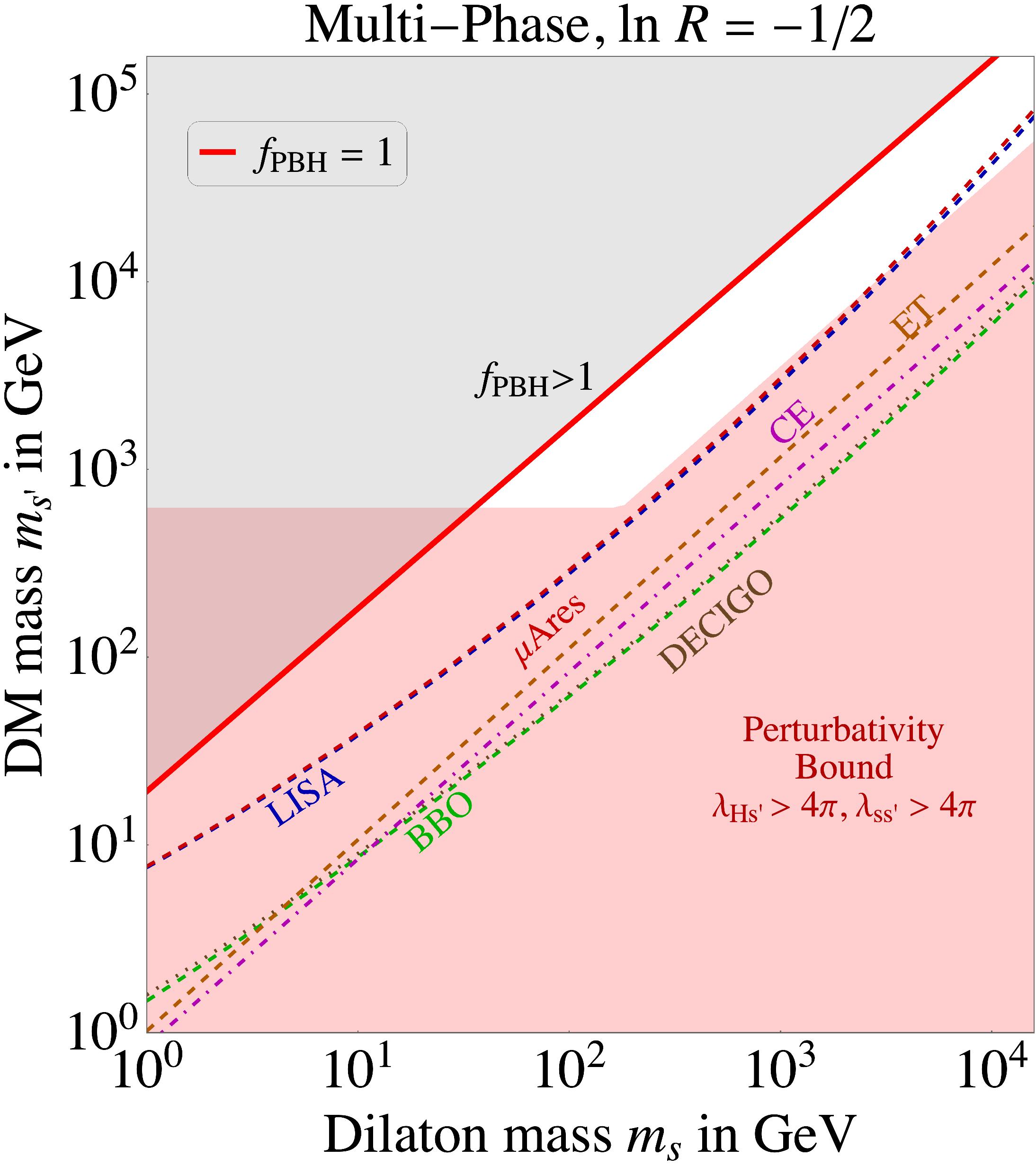}
   \qquad    \includegraphics[height=7cm,width=7cm]{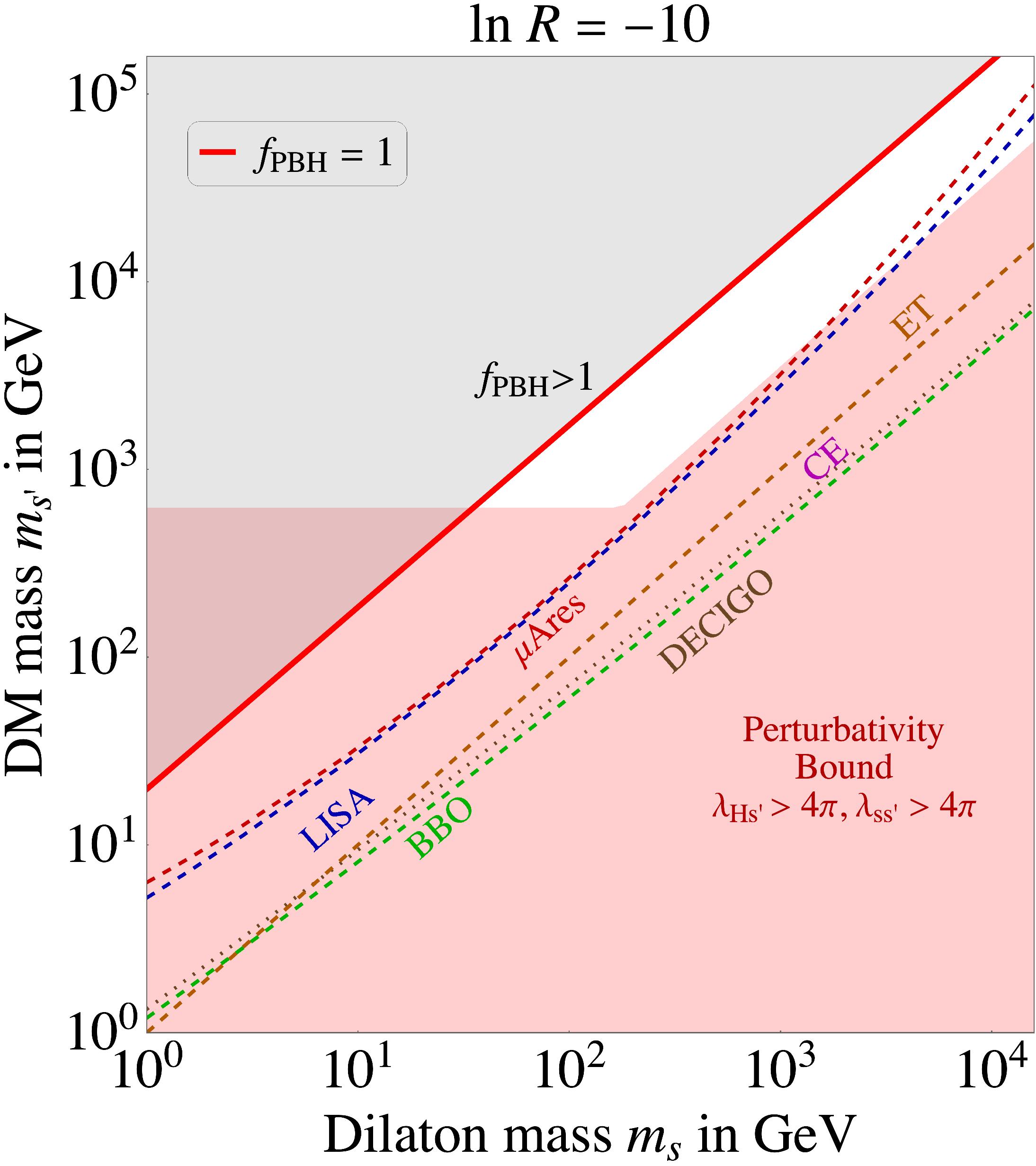}$$
\caption{\label{fig:figGW}\em GW are observable inside the bands.
Only zoomed in for LISA and CE GW detectors on the left and showing all GW detectors on the right. The arrows show the range of the parameter space that will see a signal in LISA after 4 years of running.}
\end{figure}

\begin{figure}[t]
\centering
\includegraphics[width=0.45\textwidth]{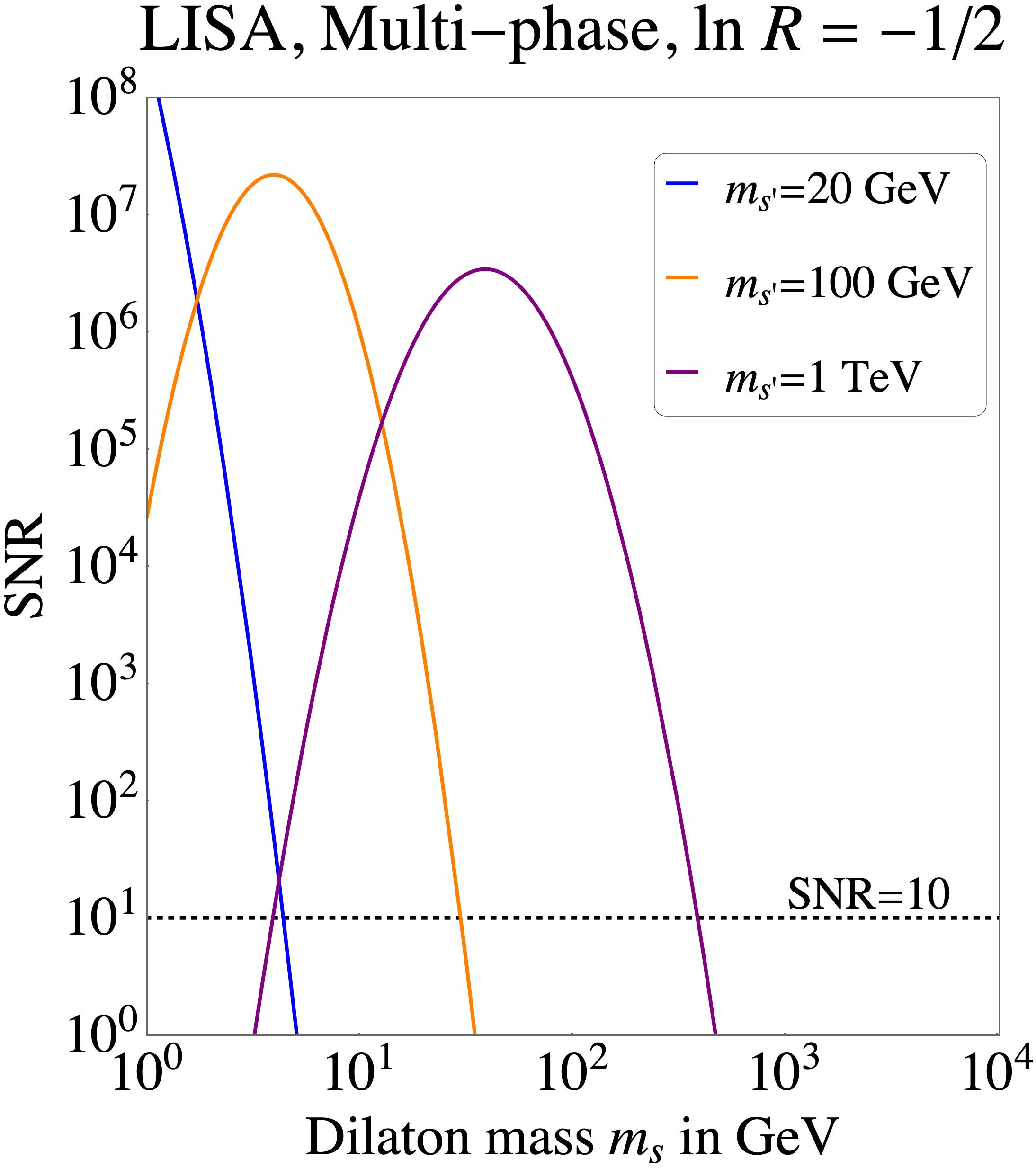}
\includegraphics[width=0.45\textwidth]{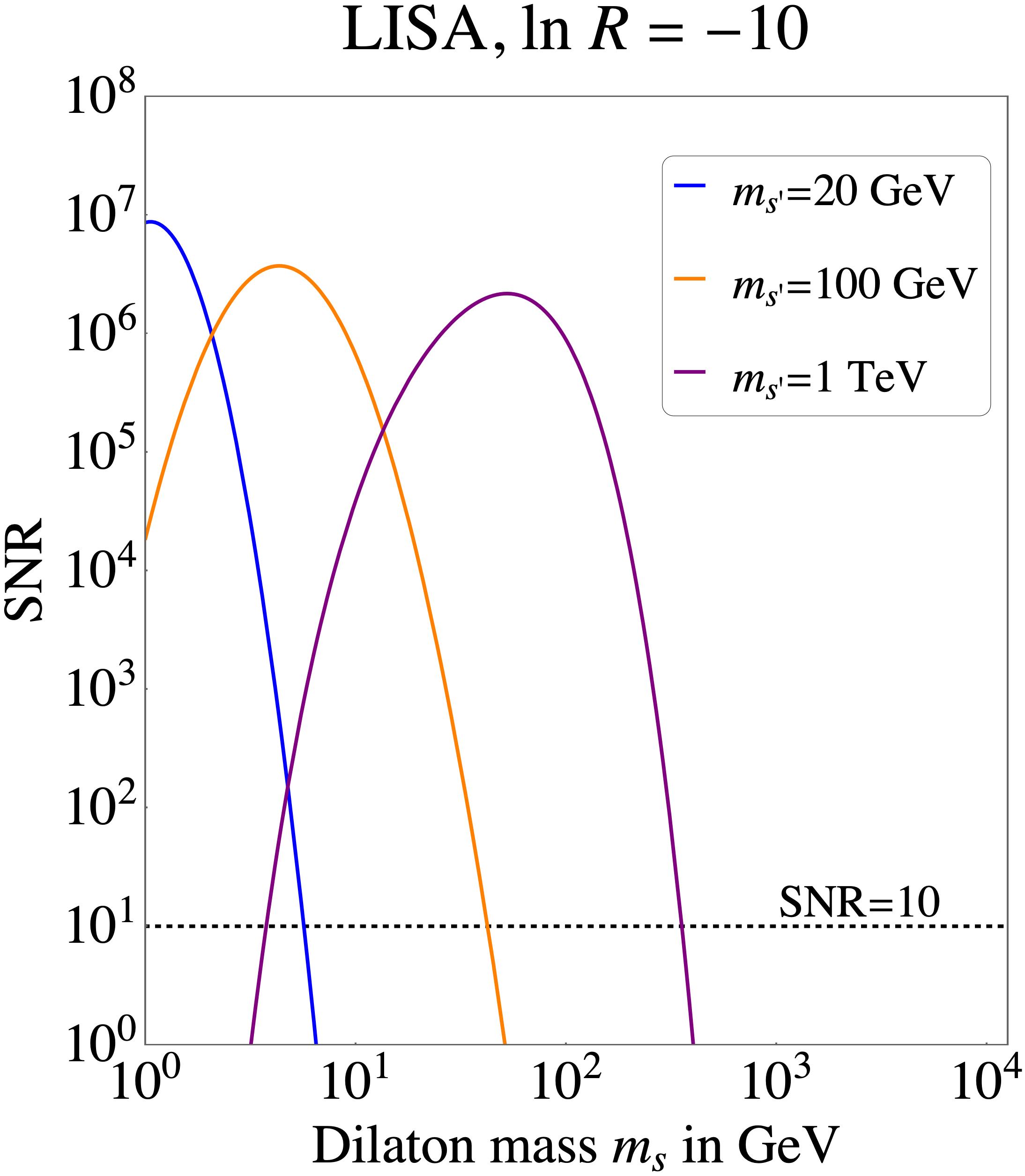}
\caption{\label{fig:LISASNR}\em Expected Signal-Noise Ratio. at LISA as function of the dilaton $m_s$ for different fixed values of the DM mass $m_{s'}$,
and neglecting astrophysical foregrounds. A SNR $\gtrsim 10$ allows GW detection.  }
\end{figure}

Fig.\fig{mainfig} also shows that significant GW signals arise in the white allowed region (see also~\cite{2408.16475}).
Each future GW experiment would probe the region in parameter space above the indicated curve,
up to the allowed boundary where $f_{\rm PBH}\approx 1$. 
We do not study the region of parameter space where QCD (rather than nucleation) ends the phase transition.
More in detail, fig.\fig{figGW} shows how the various possible GW experiments cover the parameter space,
by being sensitive to different frequencies.
As a specific example, fig.\fig{LISASNR} shows that the expected Signal-to-Noise Ratio at LISA can reach large values.
However, we here neglected the astrophysical foregrounds, 
expected with the spectrum illustrated in fig.\fig{plotGW}.
It's difficult to anticipate how much foregrounds will limit the sensitivity to primordial GW,
as some foregrounds can be partially subtracted.

\subsection{Particles as DM}
If $s'$ is stable, in most of the parameter space the DM abundance is dominated by its relic freeze-out abundance 
rather than by primordial black holes.
It is then interesting to explore the gravitational waves signals
that arise under the assumption that thermal relic particle DM makes all the cosmological DM abundance.
Since the spectrum of gravitational waves produced from bubble collisions is roughly universal, eq.\eq{GW_shape},
gravitational waves can be characterised by their peak frequency and abundance.
Fig.\fig{DMGW} shows the predictions for such two quantities
for different values of the DM mass (squared points).

\renewcommand{\topfraction}{0.7}


\section{Model with SU(2) vector DM}\label{sec:modSU2}
As a second model, we add to the SM 
a complex scalar $S$ doublet under an extra `dark' $\SU(2)_X$ gauge group with gauge coupling $g_X$~\cite{1805.01473}.
The dimension-less potential is 
\beq  V_{\rm tree} = V_\Lambda + \lambda_H |H|^4 + \lambda_S |S|^4 + \lambda_{HS} |H S|^2 .\label{eq:V}\eeq
In the unitary gauge the scalars can be expanded as
\beq S = \frac{1}{\sqrt{2}} \begin{pmatrix}
0\cr
s
\end{pmatrix},\qquad
H = \frac{1}{\sqrt{2}}
\begin{pmatrix}
0\cr h
\end{pmatrix}.\eeq
The relevant one-loop $\beta$ functions is 
\begin{equation}
\beta_{\lambda_S} \simeq  \frac{1}{(4 \pi)^2} \frac{9 \,g_X^4}{8}
\end{equation}
while, in this model, $\lambda_{HS}$ only gets multiplicatively renormalized, and can be approximated as constant.
The $\SU(2)_X$ vectors acquire a mass $M_X = g_X w/2$ and are stable DM candidates~\cite{Hambye:2013sna}.
Like in the previous model, we again use as free parameters the dilaton mass $m_s$ and the DM mass $M_X$.
The other couplings are then approximated in terms of them as~\cite{2204.01744}
\beq
g_X \simeq \frac{4\pi\sqrt{2}  m_s}{3M_X},\qquad
\lambda_{HS} \simeq -\frac{8\pi^2 m_h^2 m_s^2 }{9M_X^4},\qquad
w\simeq \frac{3 M_X^2}{2\sqrt{2}\pi m_s}.
\eeq
The mixing angle between the Higgs and the scalaron is
\beq\sin2\theta = \frac{v^2\sqrt{8\lambda_H|\lambda_{HS}|}}{m_s^2 - m_h^2}.\eeq
The $s$-wave cross-sections for DM annihilations
$XX\leftrightarrow ss$ plus  semi-annihilations $XX\leftrightarrow X s$ and for DM direct detection are~\cite{Hambye:2013sna}
\beq  \sigma_0 =\frac{65g_X^4}{6912\pi M_X^2}=\frac{260\pi^3 m_s^4}{2187 M_X^8},\qquad
\sigma_{\rm SI}\simeq    \frac{64\pi^3 f_N^2  m_N^4}{81 M_X^6}.
\eeq
The cosmological DM abundance is again reproduced when $\sigma_0 \approx 1/(23\TeV)^2$.
We restrict our analysis to the regime where $g_X\sim 1$ is large enough that super-cooling is ended by nucleation, rather than by QCD.
Then $T_{\rm RH} = (136/64 g_*)^{1/4} M_X/\pi \approx M_X/8.5$ is again larger than the DM decoupling temperature.

\begin{figure}[t]
\centering
\includegraphics[width=0.65\textwidth]{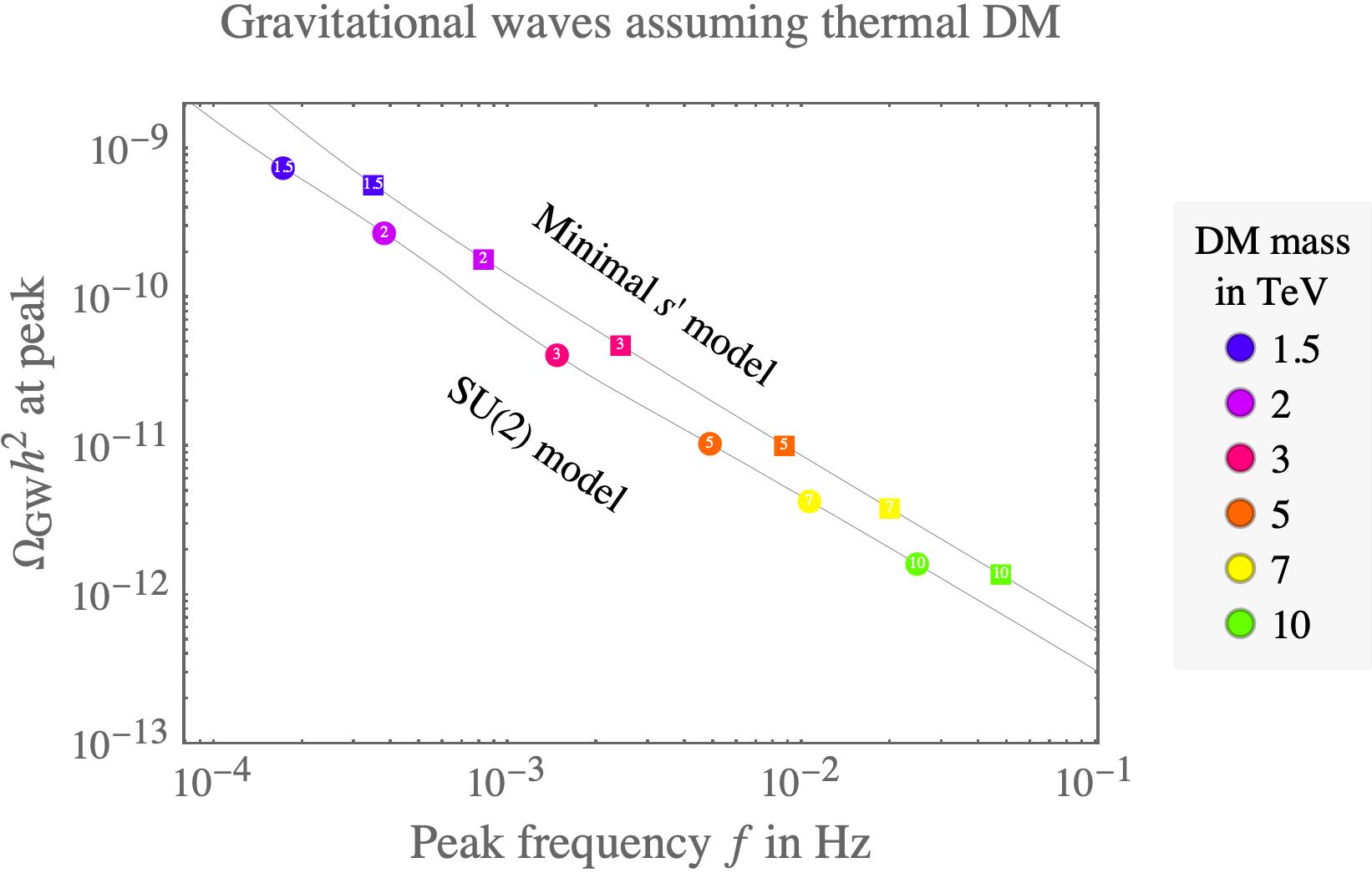}
\caption{\label{fig:DMGW}\em Predicted peak frequency and abundance of  gravitational waves 
as function of the particle DM mass, assuming that it reproduces the cosmological DM abundance, in the two models of section~\ref{sec:mods'} and~\ref{sec:modSU2}. }
\end{figure}

\begin{figure}[t]
$$\includegraphics[width=0.45\textwidth]{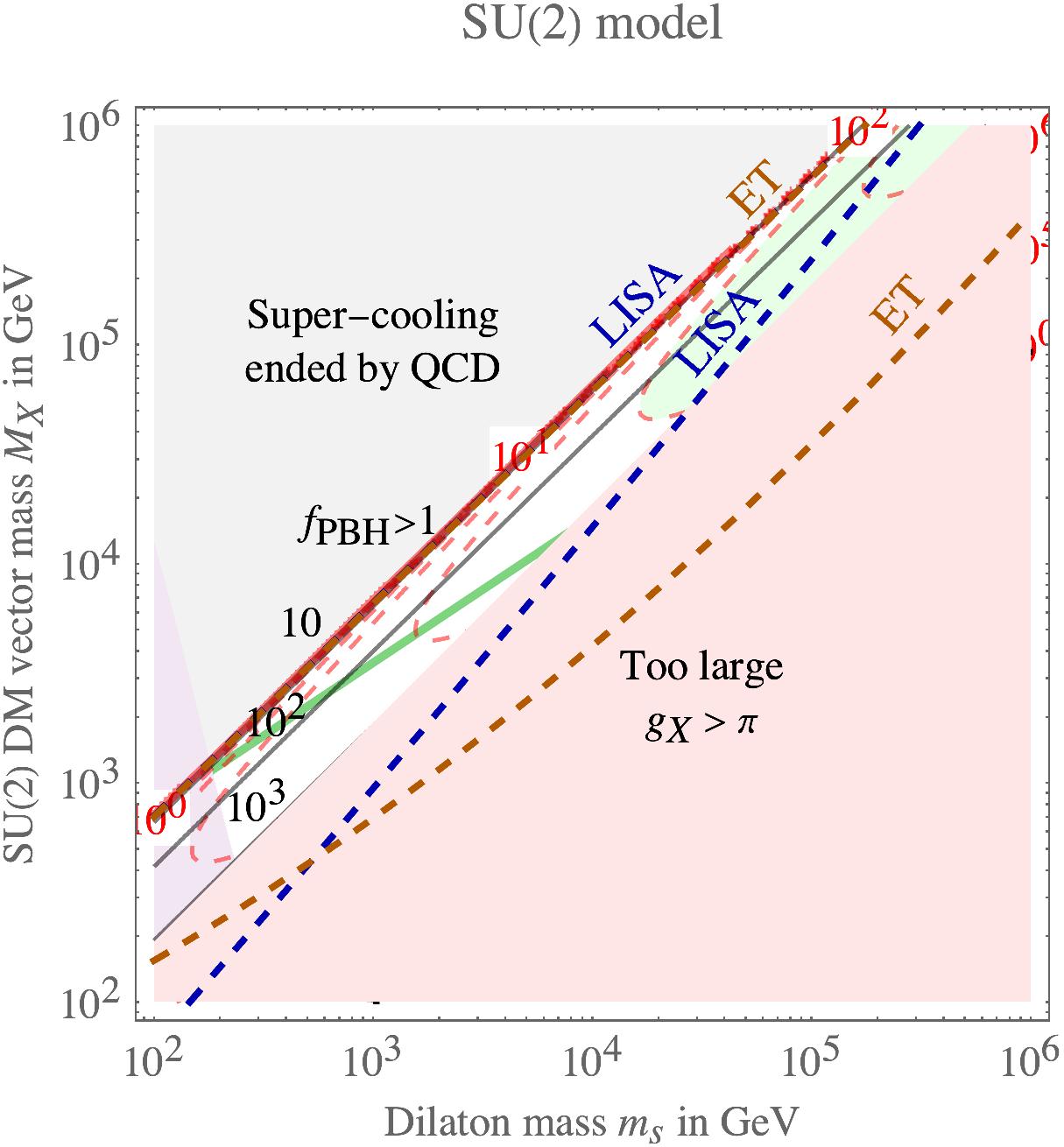}\qquad
\includegraphics[width=0.45\textwidth]{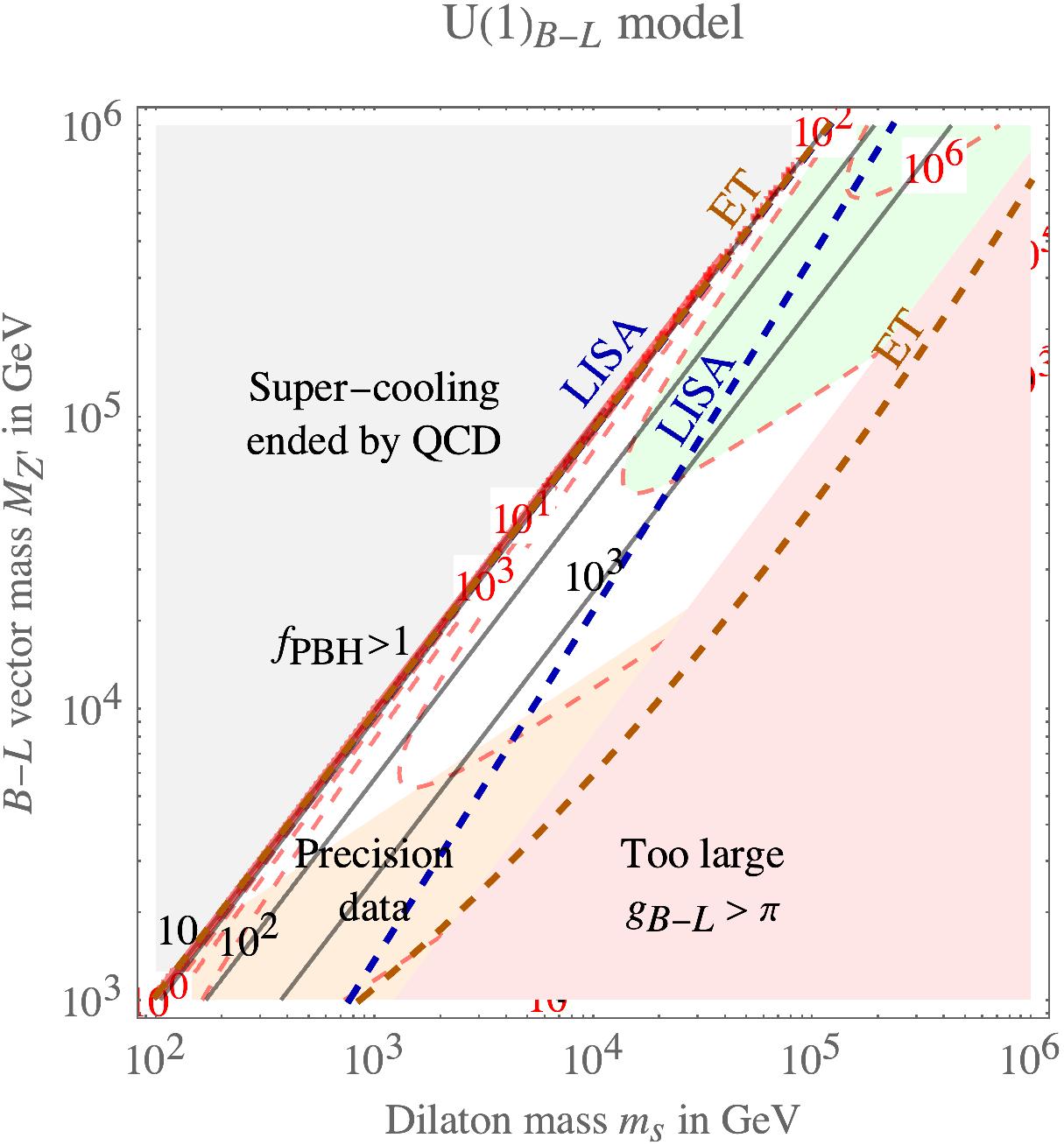}$$
\caption{\label{fig:ModelU1Tbeta}\em Summary plots for the models based on an extra dark $\SU(2)$ (left) or $\U(1)_{B-L}$ model (right).
The curves mostly have the same color coding as in the summary plot for the minimal model, fig.\fig{mainfig}
The cosmological relic abundance of particle (of Primordial Black Hole) DM is over-abundant above the green (red) curve.
The shaded regions are excluded by DM direct detection (purple);
bounds on precision data (orange);
too large quartic couplings (pink).
We omit the sub-leading bounds from $h\to ss$ and $h/s$ mixing.
We also show contour curves of $T_{\rm nuc}$ (red dashed) and of $\beta/H$ (black solid).
Stochastic gravitational waves are observed in the regions inside the dashed curves, at the indicated possible future detectors.}
\end{figure}

The above discussion shows that, up to order unity factors, the physics is qualitatively similar to the minimal singlet model of section~\ref{sec:mods'}.
So the analysis proceeds in parallel: the main results are summarised by the left panel of fig.\fig{ModelU1Tbeta}, which is similar to
fig.\fig{mainfig} for the minimal singlet model.
Predictions for gravitational waves under the assumption that particle DM matches the cosmological DM density~\cite{Hambye:2013sna,1809.01198}
are shown in fig.\fig{DMGW} (circle points).
The squared points show the predictions of the minimal $s'$ model discussed in the previous section.
The predictions of the two models are qualitatively similar, up to order order unity factors,
because different models share the following common feature:
DM induces the running and the thermal potential for 
dynamical symmetry breaking.

The phase transition in the non-scale-invariant version of the model was studied in~\cite{2408.05167}.

\section{Model with U(1)$_{B-L}$ gauge group}\label{sec:modU1}
Finally, we consider a similar model where the dark gauge group is now an Abelian U(1) factor.
Unlike in the SU(2) model, the U(1) vector is not a stable DM candidate, as long as no extra symmetry is added to forbid 
its kinetic mixing with the hyper-charge U(1)$_Y$ vector.
Furthermore, the extra U(1) needs not to be dark, 
as the SM field content allows for a possible U(1)$_{B-L}$ extra gauge symmetry.
We identify U(1) with $\U(1)_{B-L}$ and denote as $g_{B-L}$ its gauge coupling, and as $Z'$ its vector.
To break $\U(1)_{B-L}$, the scalar $S$ must have a non-vanishing charge $q_S$ under  $\U(1)_{B-L}$.
The case $q_S=2$ is often considered, such that $S$ can also provide mass to the right-handed neutrinos via a Yukawa coupling.
The following considerations largely depend only on the product $q_S g_{B-L}$.
The scalar potential is
\beq
V_{\rm tree} =V_\Lambda + \lambda_H |H|^4  +\lambda_S |S|^4   + \lambda_{HS} |HS|^2  .
\eeq
In the unitary gauge the scalars can be expanded as
\beq S = \frac{s}{\sqrt{ 2}},\qquad
H = \frac{1}{\sqrt{2}}
\begin{pmatrix}
0\cr h
\end{pmatrix}.\eeq
At one loop, the $\lambda_S$ quartic runs as  
\begin{equation}
\beta_{\lambda_S}\simeq  \frac{6 (q_S g_{B-L})^4}{(4\pi)^2}.
\end{equation}
We use as free parameters $m_s$ and $M_{Z'}$, in order to proceed in parallel with the previous models
(although $Z'$ is now not a DM candidate).
The other parameters are then approximated in terms of $m_s$ and $M_{Z'}$ as
\beq
q_S g_{B-L} \simeq \sqrt{\frac23}\ \frac{2\pi  m_s}{M_{Z'}},\qquad
\lambda_{HS} \simeq -\frac{8\pi^2 m_h^2 m_s^2 }{3M_{Z'}^4},\qquad
w\simeq \sqrt{\frac32}\frac{ M_{Z'}^2}{2\pi m_s}.
\eeq
At the minimum $\med{s}=w$ the U(1)$_{B-L}$ $Z'$ vector acquires mass
$M_{Z'}=q_S g_{B-L} w$ and decays quickly.
In this model no particle DM candidate is present, and only primordial black holes can provide DM.
Furthermore, electroweak precision data imply $M_{Z'}/g_{B-L}  \gtrsim 7\TeV$~\cite{hep-ph/0604111,0909.1320} up to corrections due to kinetic mixing with hypercharge.
The re-heating temperature is
$T_{\rm RH}=T_{\rm infl}= M_{Z'} (45 /64\pi^4 g_*)^{1/4} \approx M_{Z'}/11$.
The  $s$ thermal potential is $V_T \approx 3T^4\, J_B(q^2_S g^2_{B-L}s^2/T^2) /2\pi^2$, leading in eq.\eq{VTapprox}
to the thermal mass $m= q_S g_{B-L}  T/2$,
and to the cubic $k= 3 (q_S g_{B-L})^3 T/4\pi$.
The right-handed panel of fig.\fig{ModelU1Tbeta} summarises our results for this model.
PBH have the DM abundance along the red curve, and their mass falls in the region where PBH can be DM in the green region.
This is compatible with $T_{\rm nuc} \gtrsim \GeV$, such that super-cooling is ended by nucleation.
Again, gravitational waves arise at a detectable $\Omega_{\rm GW}\sim 10^{-8}$ level.
Such conclusions agree with a previous study~\cite{Gouttenoire:2023pxh}.

\section{Discussion and Conclusion}\label{concl}
We investigated various models of dynamical breaking of the electroweak symmetry.
A common feature is that an extra scalar $s$, neutral under the SM gauge group, is added to the Higgs sector.
Some extra particle is needed to mediate quantum corrections to the $s$ potential such that $s$ acquires
a vacuum expectation value. Different models introduce different particles:
the minimal model employs one extra scalar $s'$; other models employ vectors of an extended
U(1) or SU(2) gauge group.

In all cases the cosmological phase transition is strongly first order, leading to Gravitational Waves 
(at a level detectable by realistic future observatories)
and to Primordial Black Holes
(at a level that, in a part of the parameter space, can account for the Dark Matter relic density of the universe). 
The PBH mass depends on the parameters of the model and can fall in the range where PBH can be all of DM;
this parameter space can be tested by GW signals.

In the minimal model, this is illustrated by the examples in fig.\fig{plotPBH}.
Fig.\fig{plotGW} shows that such a range leads to detectable GW with $\mu$Hz-mHz frequencies,
while the new $s$ and $s'$ particles are too heavy to give signals at current particle colliders.
The model already contains one possible DM candidate, as the scalar $s'$ can be stable.
In this case, the $s'$ thermal freeze-out relic abundance dominates over the black hole abundance.
The regime of particle DM is realized when particles have lighter TeV-scale mass,
leading to collider signals mostly from $s$/Higgs mixing, direct detection signals, and GW with lower frequency in the $\mu$Hz-nHz range.

A similar situation is encountered in the SU(2) model, where the SU(2) vectors are stable DM candidates.

The vector  present in the U(1) model can instead decay via a kinetic mixing with hyper-charge
and/or because SM fermions are charged under it (the U(1) can be identified as U(1)$_{B-L}$).
Primordial black holes can have appropriate mass and abundance to explain Dark Matter.

\medskip

Primordial black holes  and gravitational waves in the same frequency range
can also be generated by different theories, where inflation produces large curvature perturbations at small scales, 
see e.g.~\cite{2109.01398}.
The GW spectral shapes arising from second-order tensor perturbations are different from those produced due to bubble collisions during phase transitions.

\small

\section*{Acknowledgement}
We thank Luca Marzola for collaboration and discussion during the initial stage of the project,
Marcos Flores, Marek Lewicki and Tomer Volansky for discussions.

\medskip 

\footnotesize


\end{document}